# Defect Diagnosis in Rotors Systems by Vibrations Data Collectors Using Trending Software

Hisham A. H. Al-Khazali[1]
Institute School of Mechanical & Automotive Engineering,
Kingston University,
London, SW15 3DW, UK.

Mohamad R. Askari[2]
Institute School of Aerospace & Aircraft Engineering,
Kingston University,
London, SW15 3DW, UK.

***Abstract*-Vibration measurements have been used to reliably diagnose performance problems in machinery and related mechanical products. A vibration data collector can be used effectively to measure and analyze the machinery vibration content in gearboxes, engines, turbines, fans, compressors, pumps and bearings. Ideally, a machine will have little or no vibration, indicating that the rotating components are appropriately balanced, aligned, and well maintained. Quick analysis and assessment of the vibration content can lead to fault diagnosis and prognosis of a machine's ability to continue running. The aim of this research used vibration measurements to pinpoint mechanical defects such as (unbalance, misalignment, resonance, and part loosening), consequently diagnosis all necessary process for engineers and technicians who desire to understand the vibration that exists in structures and machines.**

## I. Introduction

### A. Vibration Analysis Software

In the past, diagnosis of equipment problems using vibration analysis was mostly dependent on the ability of the maintenance technician or the plant engineer [1, 2]. However, today's vibration analysis equipment utilizes software that has greatly enhanced the analysis of vibration measurements and the prediction of equipment performance [3, 4]. Before computer-based data collectors, most vibration programs consisted of recording overall velocity, displacement, and acceleration measurements on a clipboard [5, 6]. Those with IRD instruments had a mysterious measurement available called Spike Energy [7, 8]. The measurements were transferred to charts for trending. Obviously, this was very labour intensive, but it was successful [9, 10]. Those fortunate enough to have spectrum analyzers with plotters, would take spectra and paste them to a notebook with the overall trends [11,12& 13]. This whole process of manual storage and trending of overall and spectral data was not very cost effective, but many of those programs were successful [3].

The advent of computer-based data collectors and trending software made this whole trending process much more cost-effective. The first system only recorded overall measurements [14,15& 16]. Spectral data still had to be taken with a spectrum analyzer. Those programs based only on overall measurements were usually successful in identifying a machine developing a problem [3, 17].

Vibration analysis software is an essential tool for professionals who are analyzing the vibratory behaviour of structures and machines. There are two types of vibration analysis investigations that are commonly done today [4, 18]. The first involves analyzing the mechanical vibration of new products that are being designed or tested. The second involves analyzing the vibration that exists in rotating machinery such as compressors, turbines, and motors [19, 20]. Advances in computing power and software have greatly enhanced the vibration analysis process for both new and existing products. The most common software package for analyzing new products is finite element analysis software. Vibration in existing machinery is analyzed with integrated software that enhances the vibration analysis equipment [14, 21].

### B. Vibration Testing

Vibration testing is used to determine how well a product will withstand its expected service and transportation environments. Equipment that must withstand vibration testing includes automotive, aerospace, machinery, electrical, medical and power [22].

Performing a vibration test reproduces one of the most severe real-world environmental conditions that equipment will encounter. Since vibration testing is crucial during product development, selecting and using vibration testing equipment is an important step for engineers and product managers [23, 24].

## II. Methods And Experimental Procedures

### A. Vibration Testing Equipment

Vibration testing equipment includes accelerometers, controllers, analyzers, amplifiers, shakers, and vibration test fixtures. While each industry utilizes vibration testing in a unique manner, the most important components of a vibration test system are basically the same [25,26& 27]. Vibration is measured and controlled using displacement transducers or accelerometers. Then, do the experimental testing using the electromagnetic shaker test, installed two accelerometers (model 339A32, SN 4851, sensitivity 96.5 & 99.6 mV/g) in Y& Z direction see picture (1), it was attached to the test structure with creating a computer when taking readings in file that was dimensions and introducing it with the data within the program (smart office). Vibration monitoring equipment includes PC-based controllers and analyzers. Vibration is transmitted to the test product using stiff, lightweight test fixtures. Vibration shakers are available in electrodynamics or hydraulic versions. Electrodynamics' shakers are used for smaller products that require smaller displacements and a larger frequency range. It was powered by amplifiers that may rival a radio station for electrical power output [26, 28].

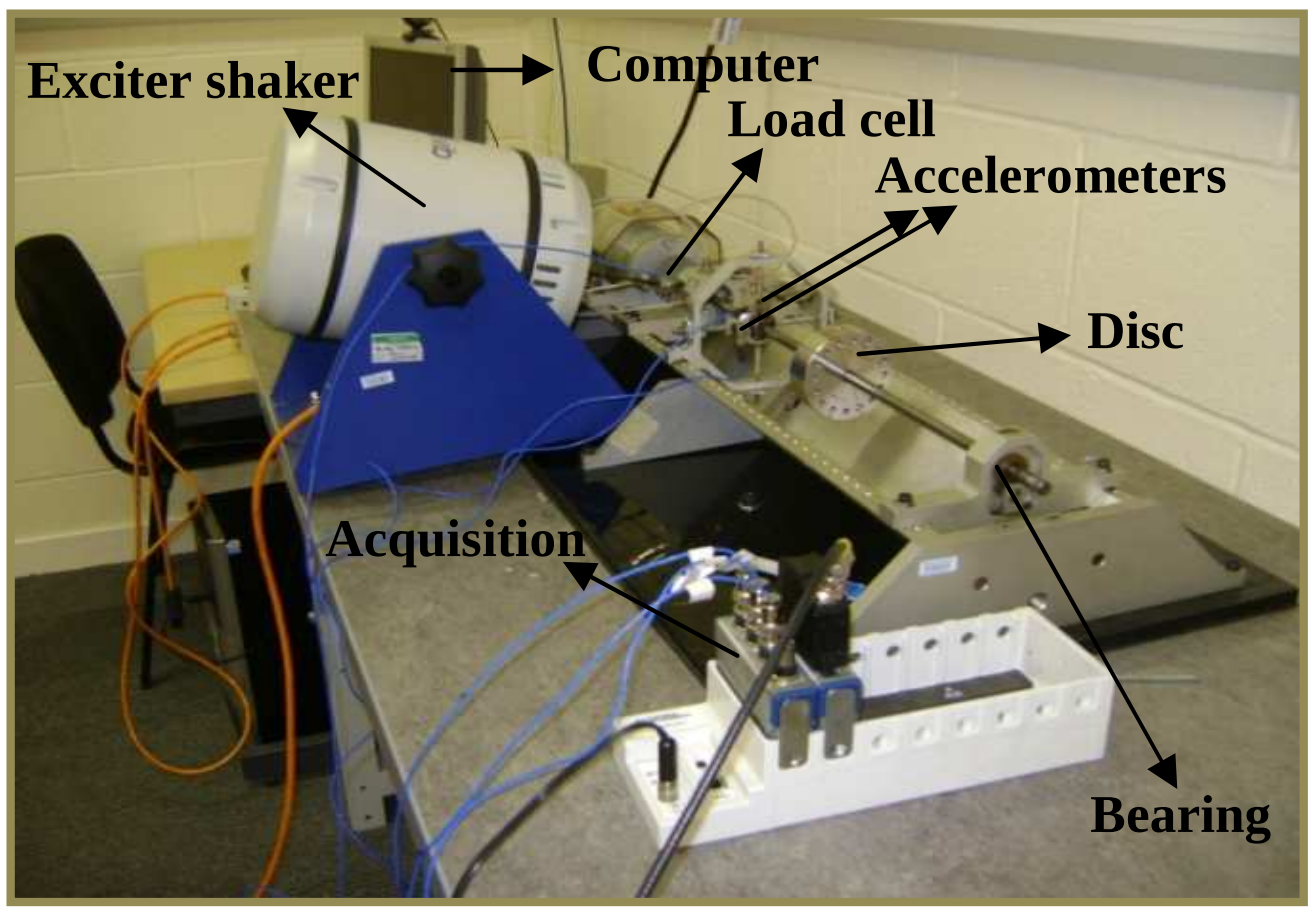

A. Rotor Rig setup for the modal testing.

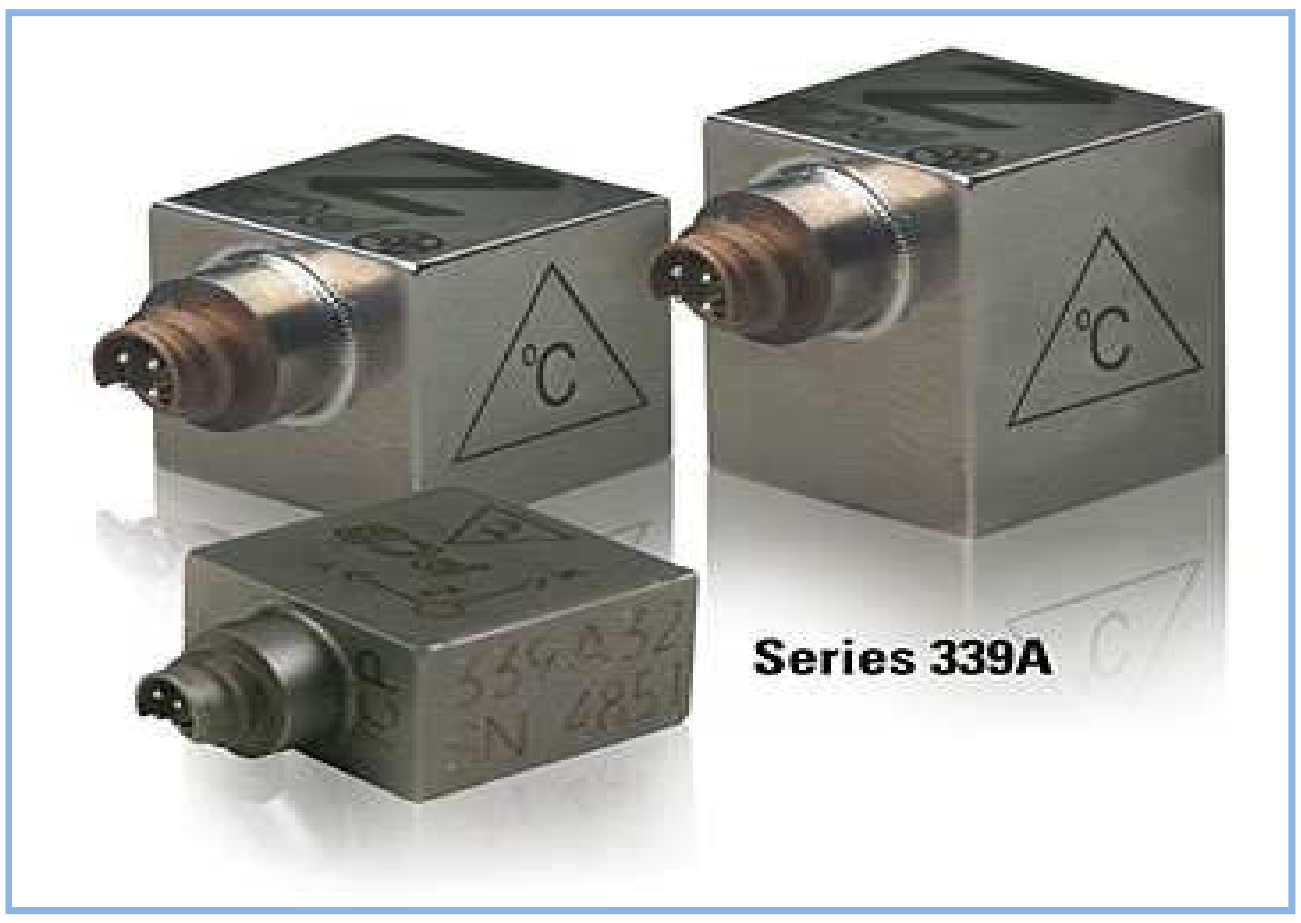

B. Different types of accelerometers, 2012.

Picture 1. Arrangement of the experiment;

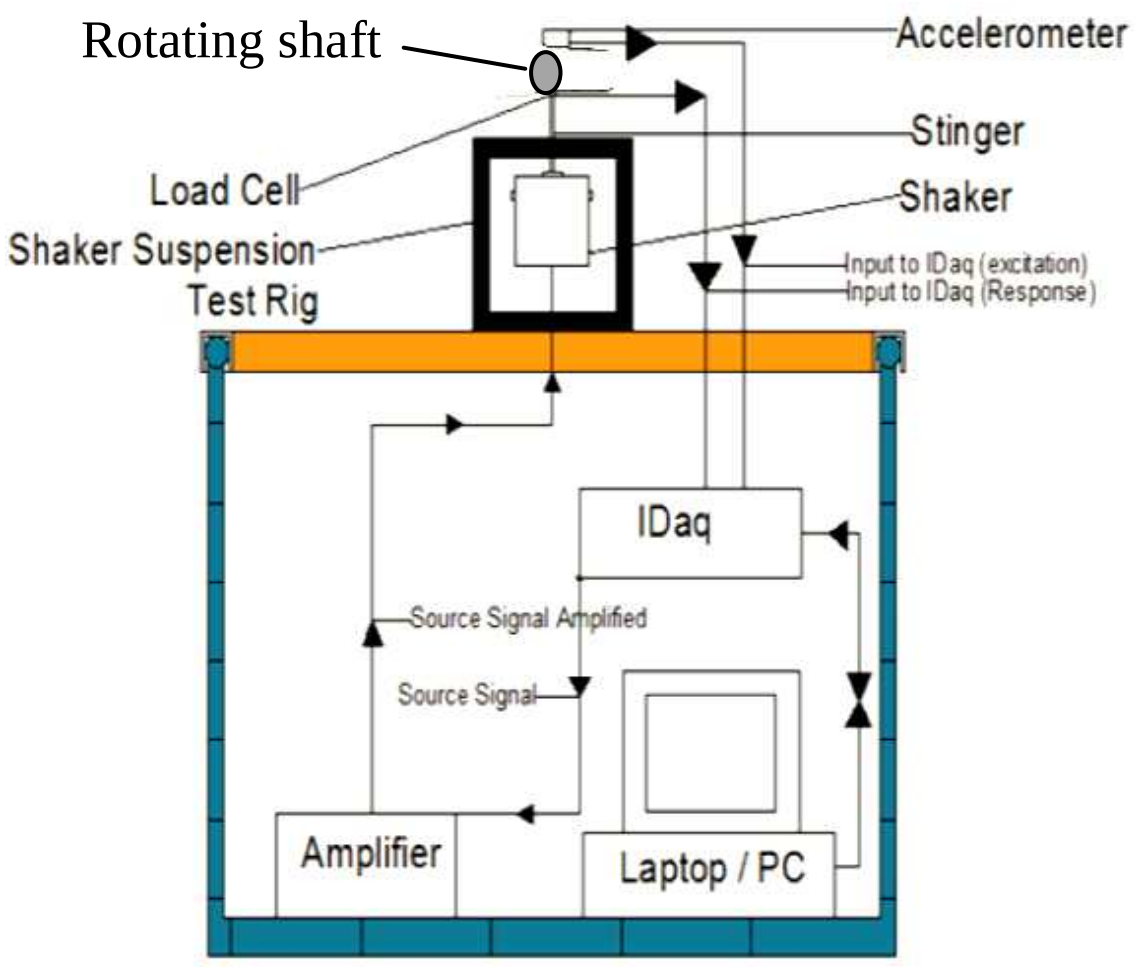

A . The conceptual design show method of the outcome results using electromagnetic Shaker test.

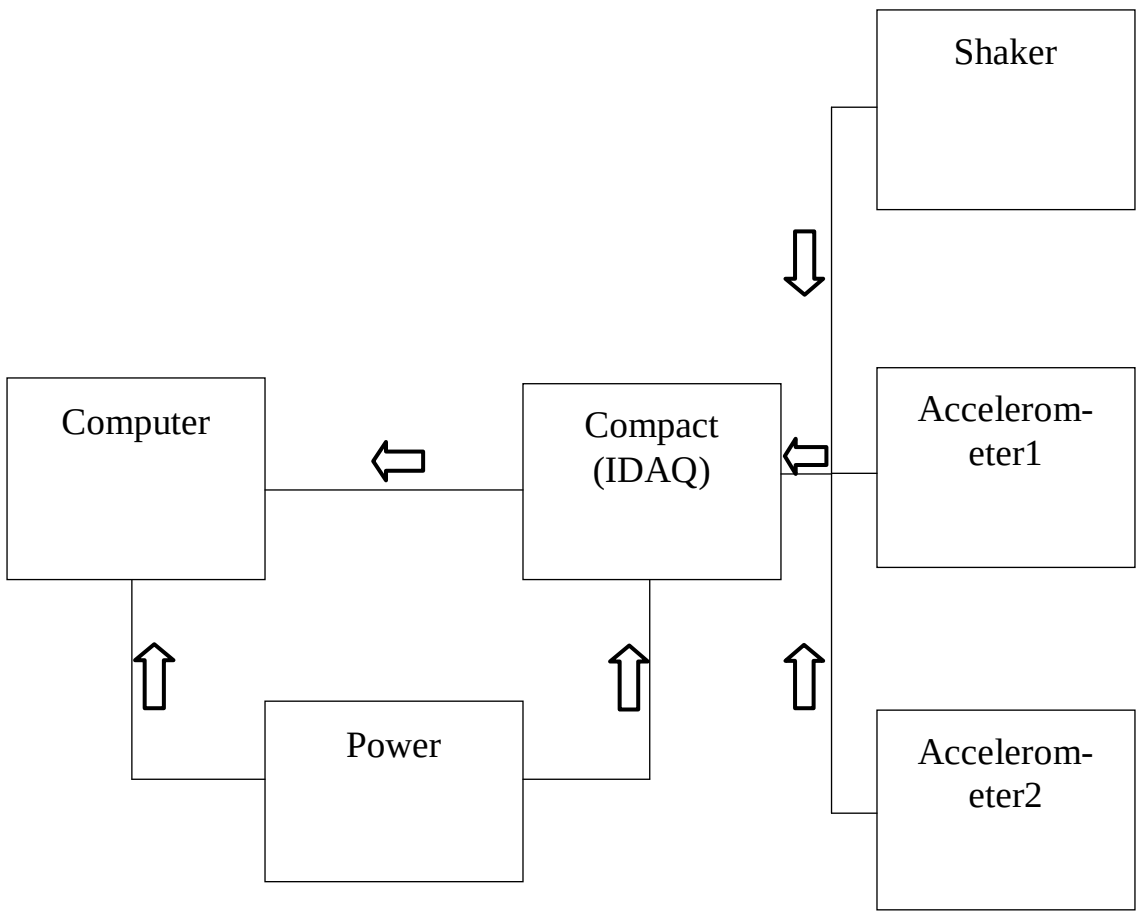

B. Block diagram of test processing controller.

Scheme 1. Test setup;

## *B. Mathematical Models*

(i) Equation of motion in rotor system for (passive) structures [23, 29], (no self-excitation).

$$[M]\{\ddot{q}(t)\}+[(C]\{\dot{q}(t)\}+[K]\{q(t)\}=\{F(t)\} \qquad (1)$$

All the above matrices are normally symmetric, after solved it we get:

$$\alpha_{jk}(\omega)=\sum_{r=1}^{n}\frac{\varphi_{jr}\,\varphi_{kr}}{\boldsymbol{M}_r\left[\omega_r^2-\omega^2+2\,i\,\omega\,\xi_r\,\omega_r\right]} \qquad (2)$$

(ii) For (active) rotating structures [17,23& 30], includes the gyroscopic effect [*G*], however in machinery where there are rotating parts as well as the supporting structure, the matrices are generally non-symmetric, and therefore need to be treated differently.

$$[M]\{\ddot{q}(t)\}+[(C+G)(\Omega)]\{\dot{q}(t)\}+[K(\Omega)]\{q(t)\}=\{F(t)\} \qquad (3)$$

The non-homogeneous part of the above equation 3, may be solved for free vibration analysis, we obtain

$$\alpha_{jk}(\omega)=\sum_{r=1}^{n}\left(\frac{2\omega_r\left(\zeta_r \operatorname{Re}\left({}_rG_{jk}\right)-\sqrt{1-\zeta_r^{2}}\left(\operatorname{Im}\left({}_rG_{jk}\right)\right)\right)+i\left(2\omega \operatorname{Re}\left({}_rG_{jk}\right)\right)}{\left(\omega_r^{2}-\omega^{2}+2i\omega\omega_r\zeta_r\right)}\right) \quad (4)$$

Equation 4, represents the receptance between two coordinates j and k for a system with n degree of freedom. The denominator is identical to the denominator of the receptance expression for an n degree of freedom system with symmetric system matrices (equation 2), but the numerator is very different.

## III. RESULTS

### A. *The Orbit Analysis Postprocessing Wizard*

This orbit analysis tool has been designed for bearing and shaft analysis. It post-processes time history file that must include a once-per-rev key-phasor signal to provide a speed and reference for top dead centre (TDC). Various filters are available to smooth the data over a number of revolutions [4, 10].

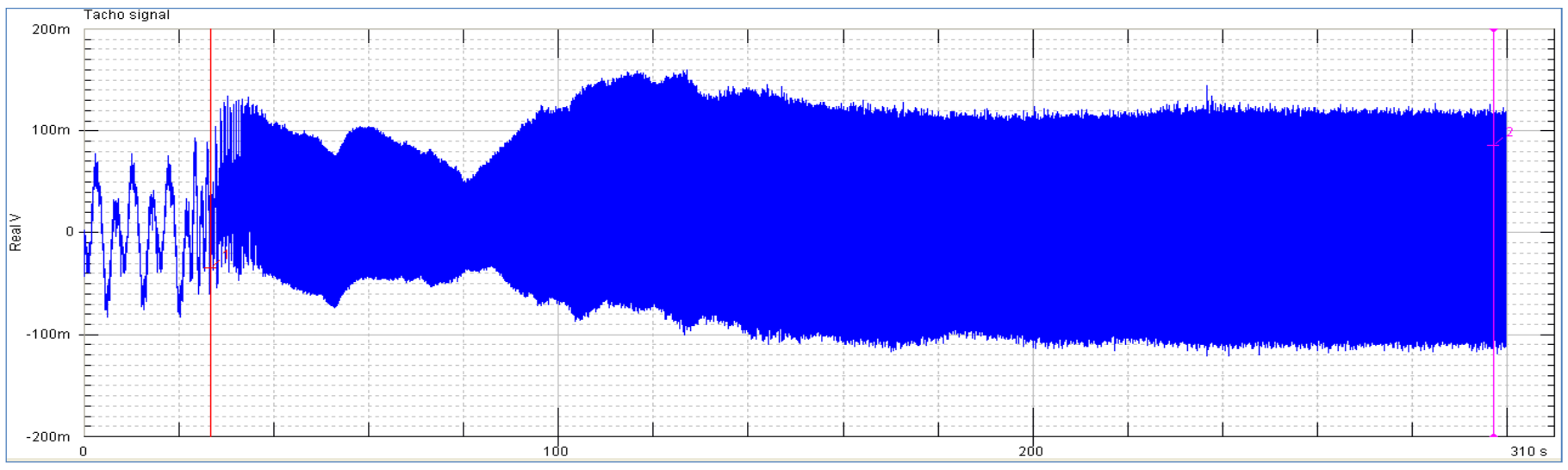

A. Real versus time (sec.)

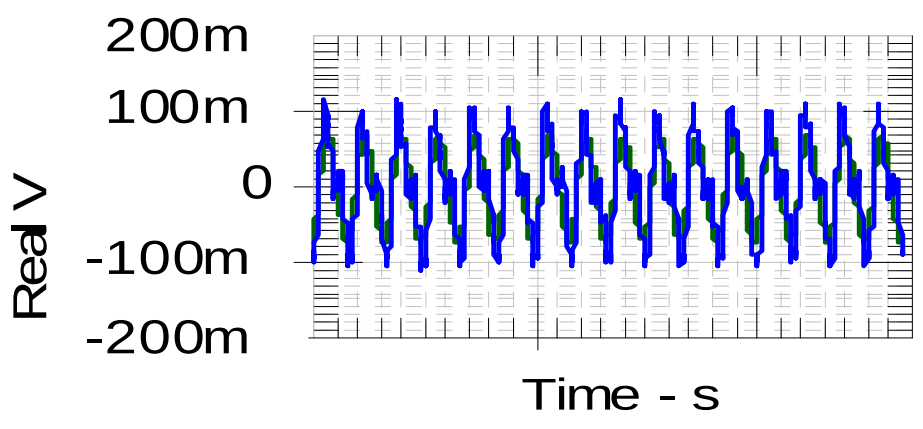

B- y, axis measurement.

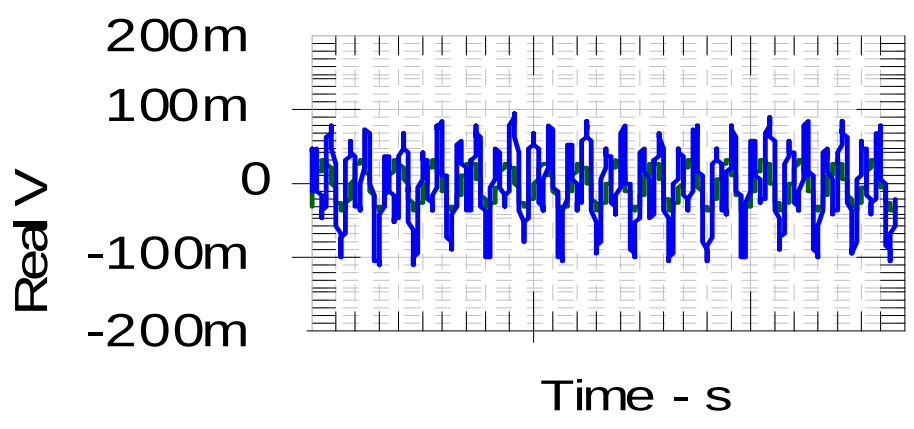

C- z, axis measurement.

Figure 1. Signal detection in (y & z) axis measurement direction;

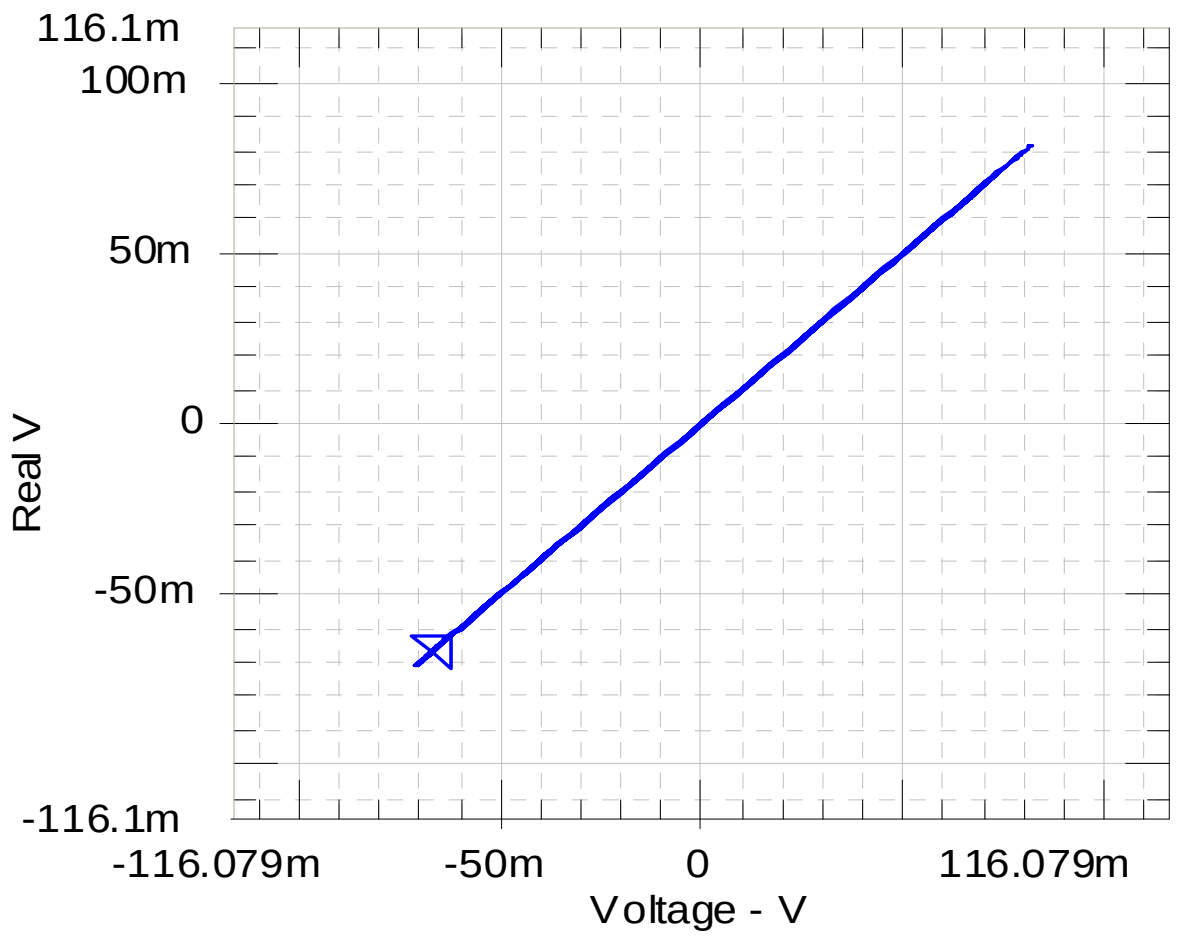

A-

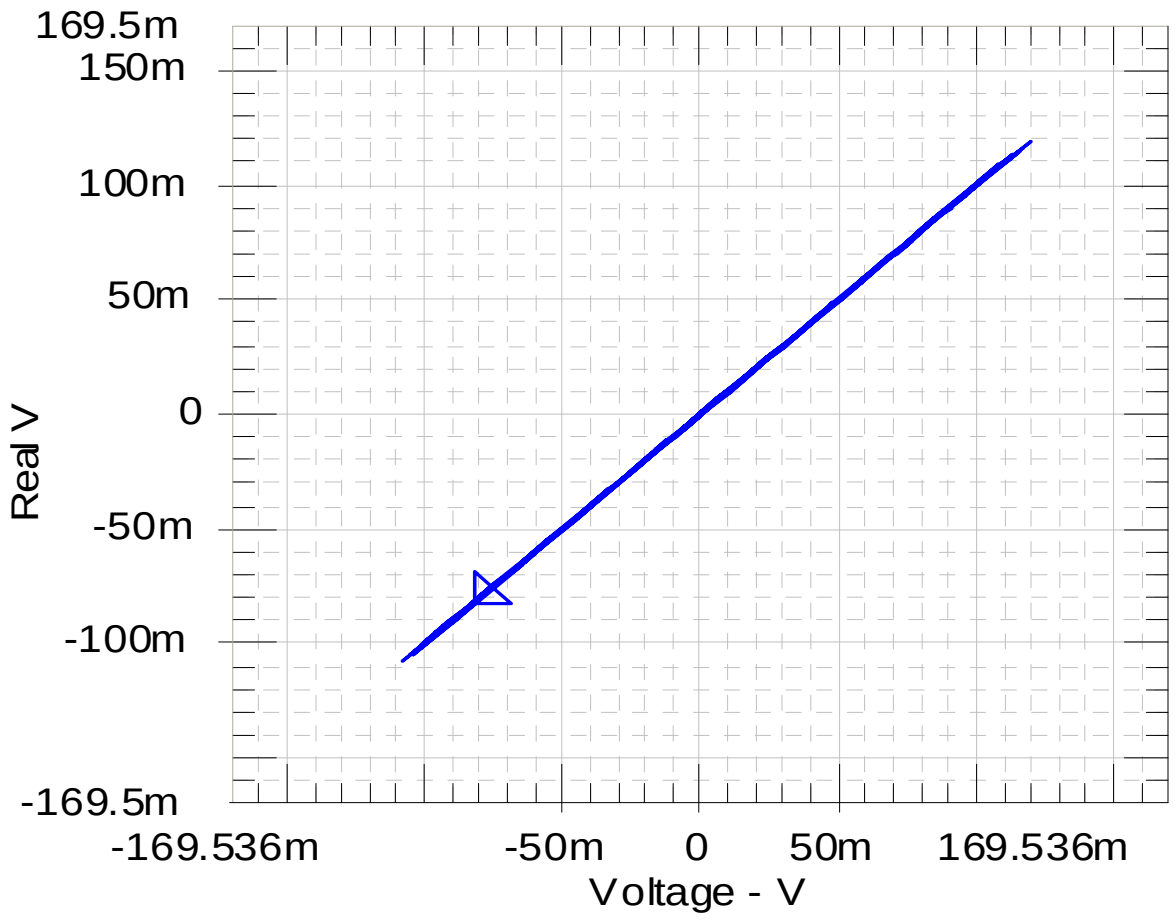

B-

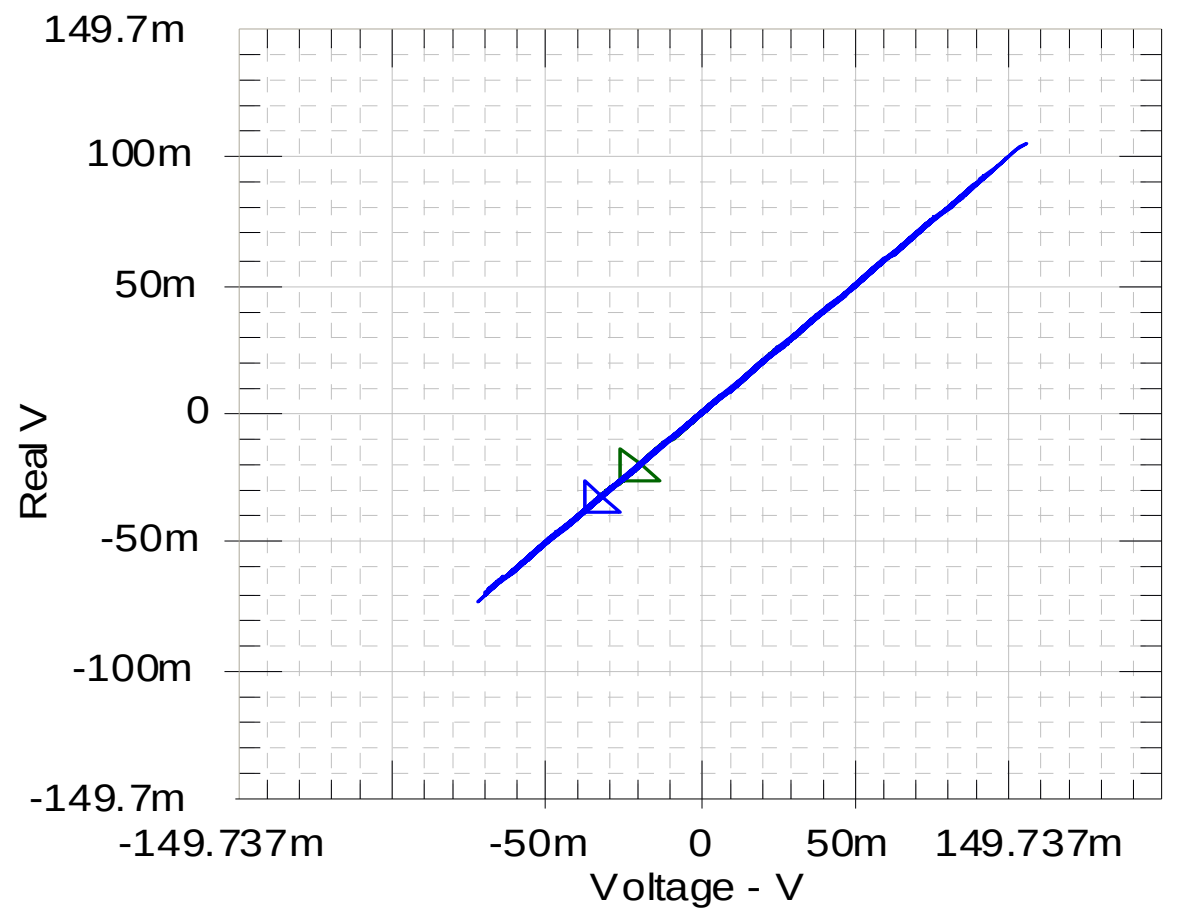

C-

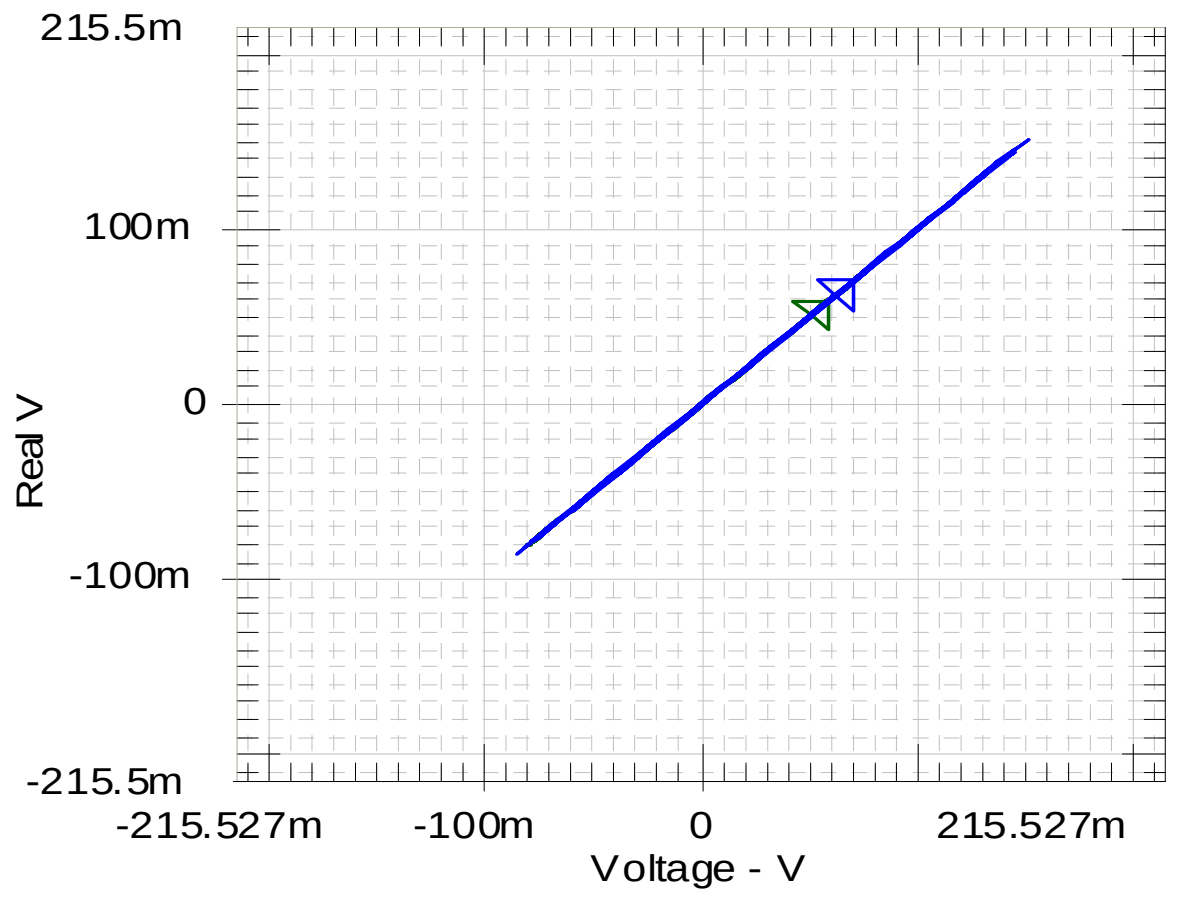

D-

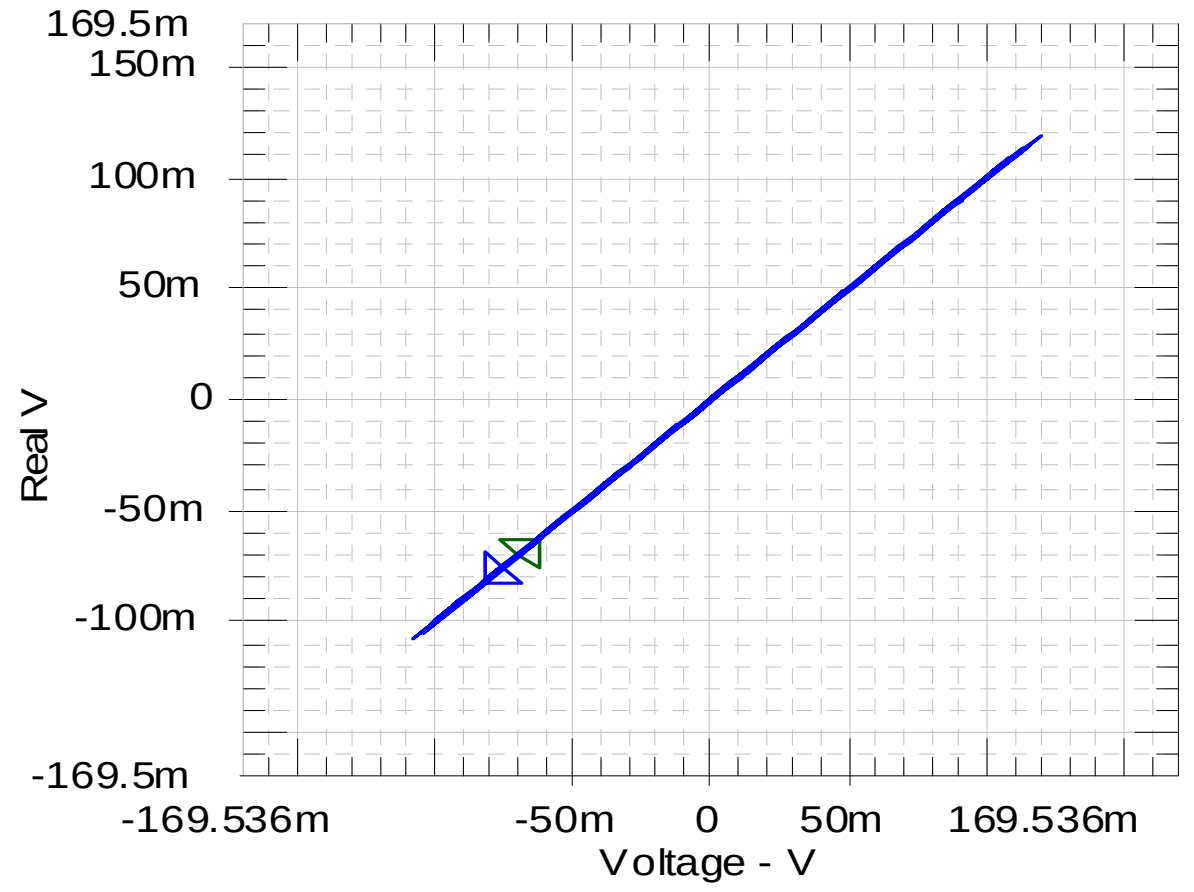

E-

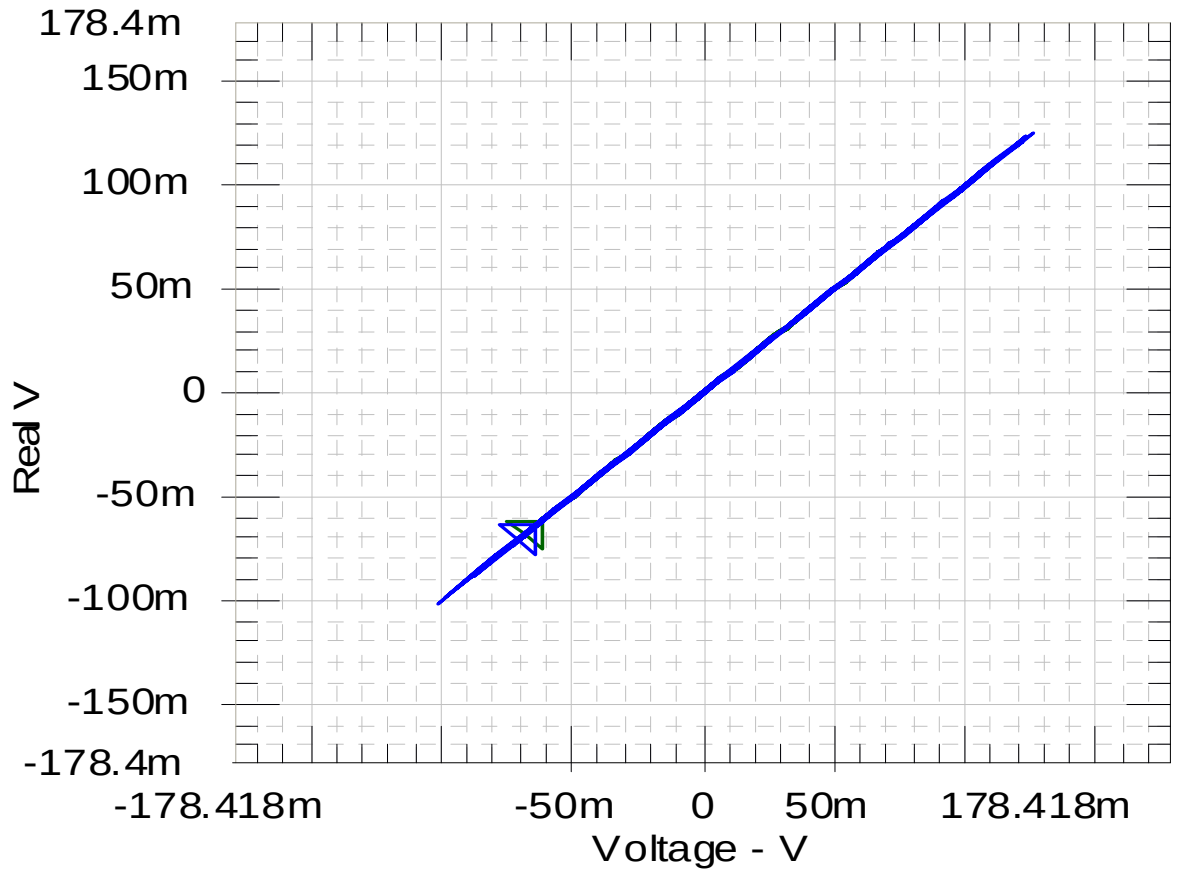

F-

Figure 2. Experimental orbit analysis direction in rotor dynamics;

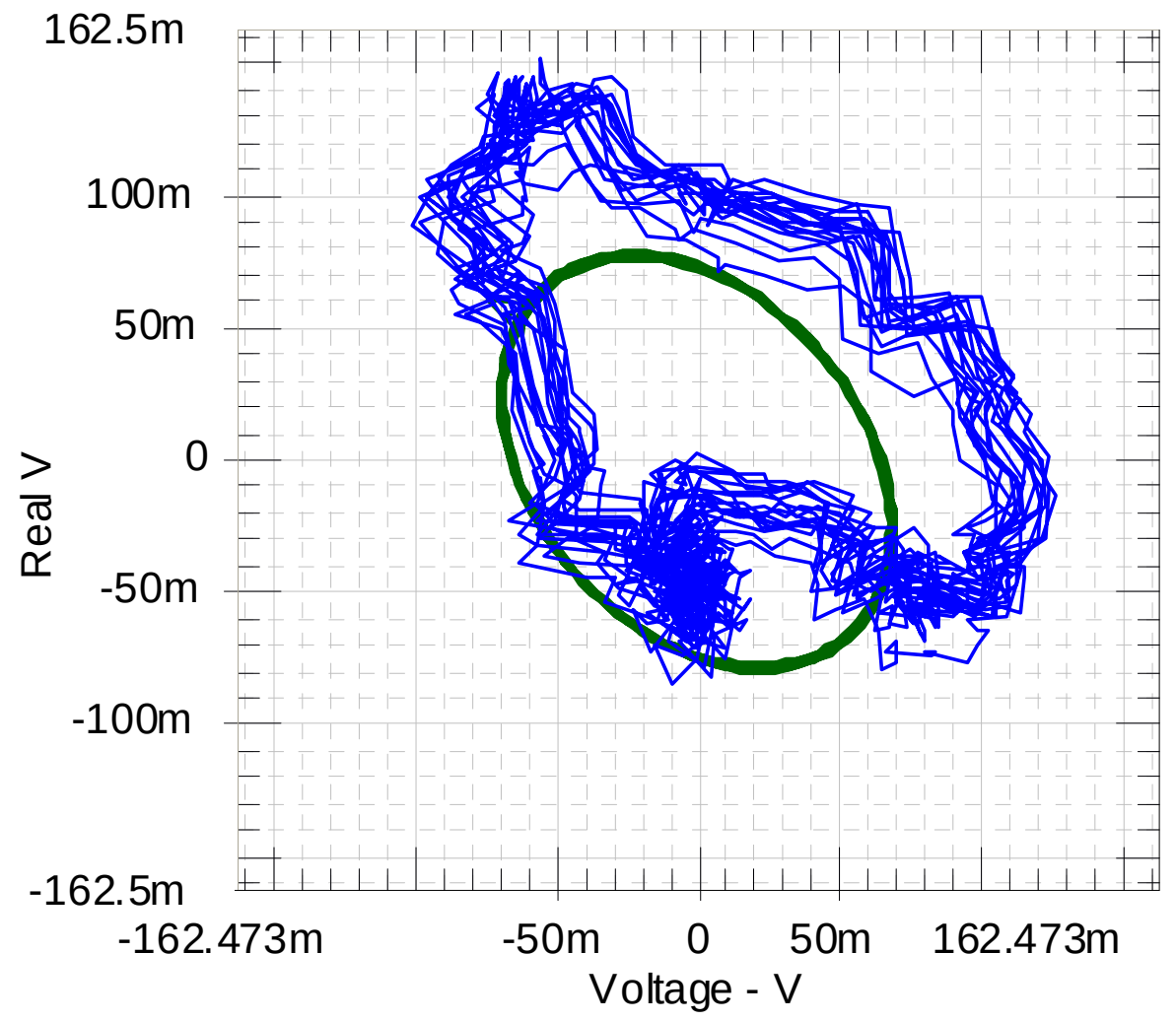

A-

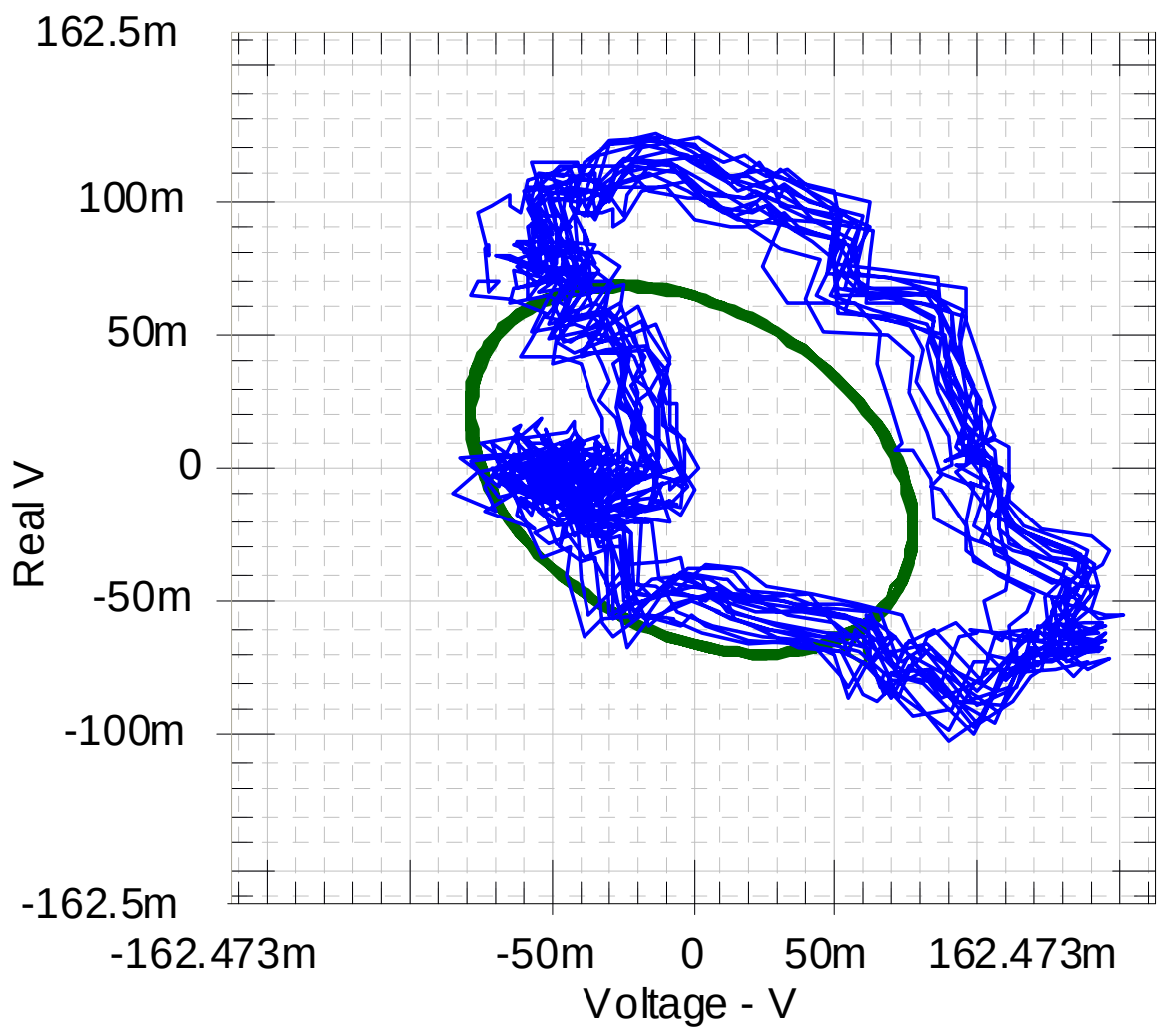

B-

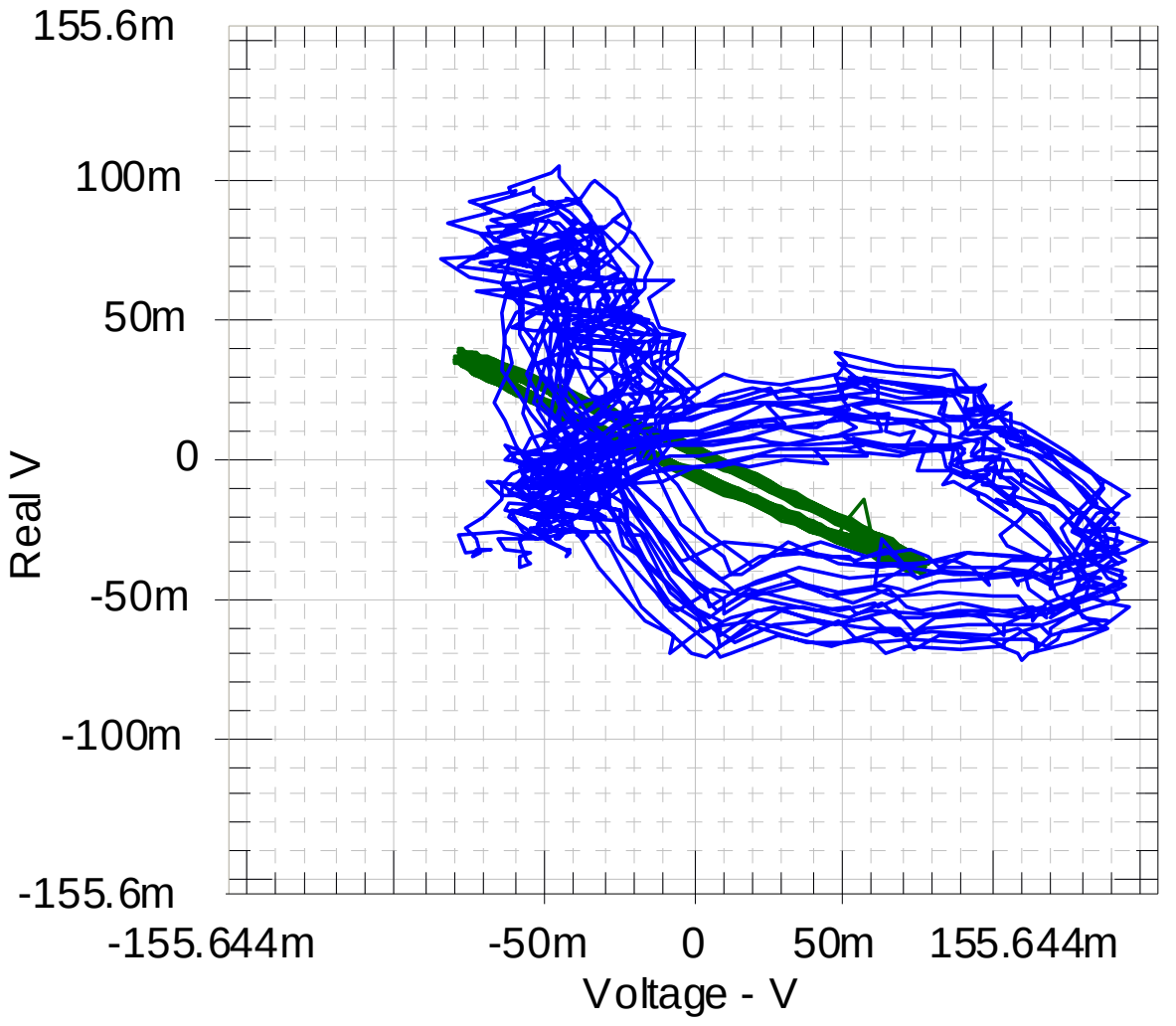

C-

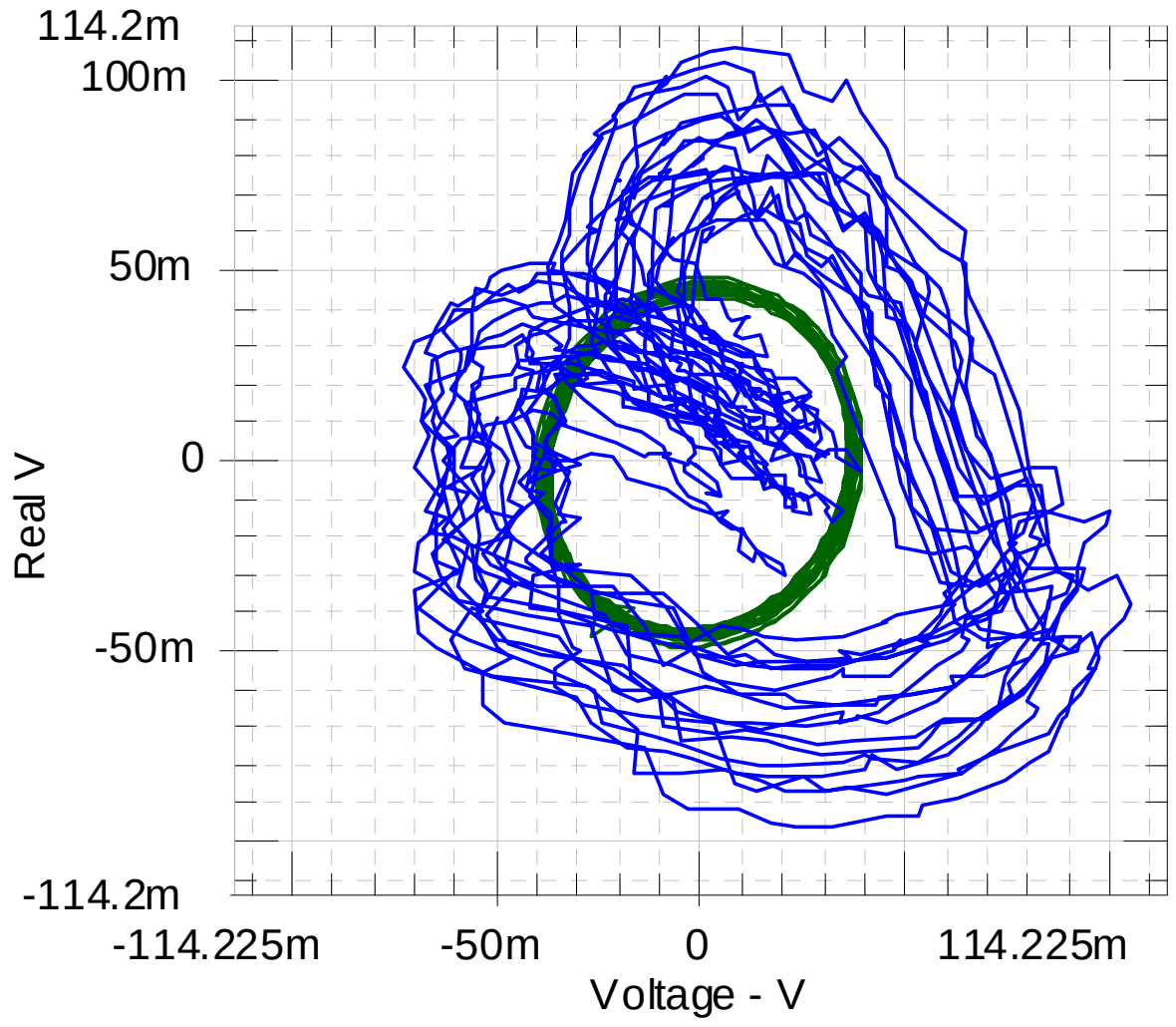

D-

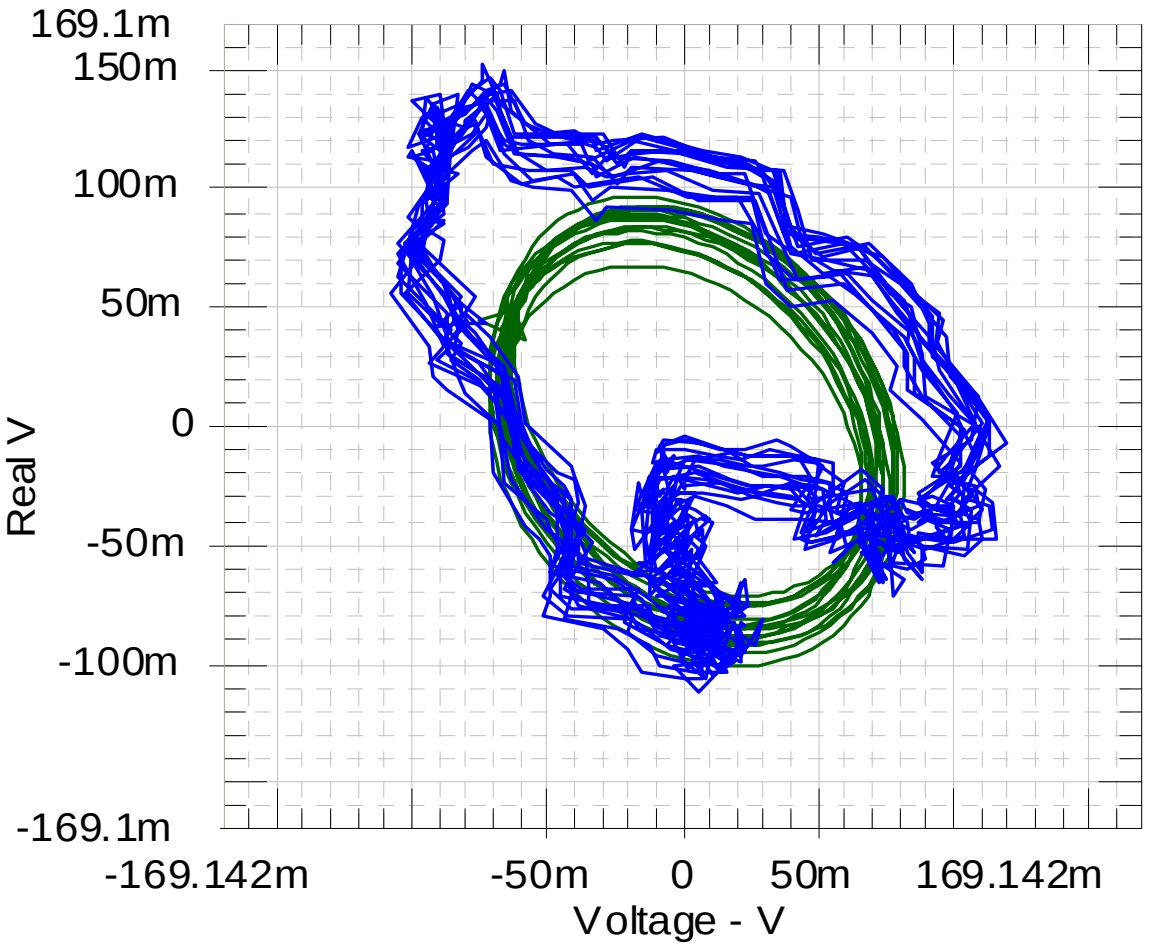

E- Without filter (HP filter).

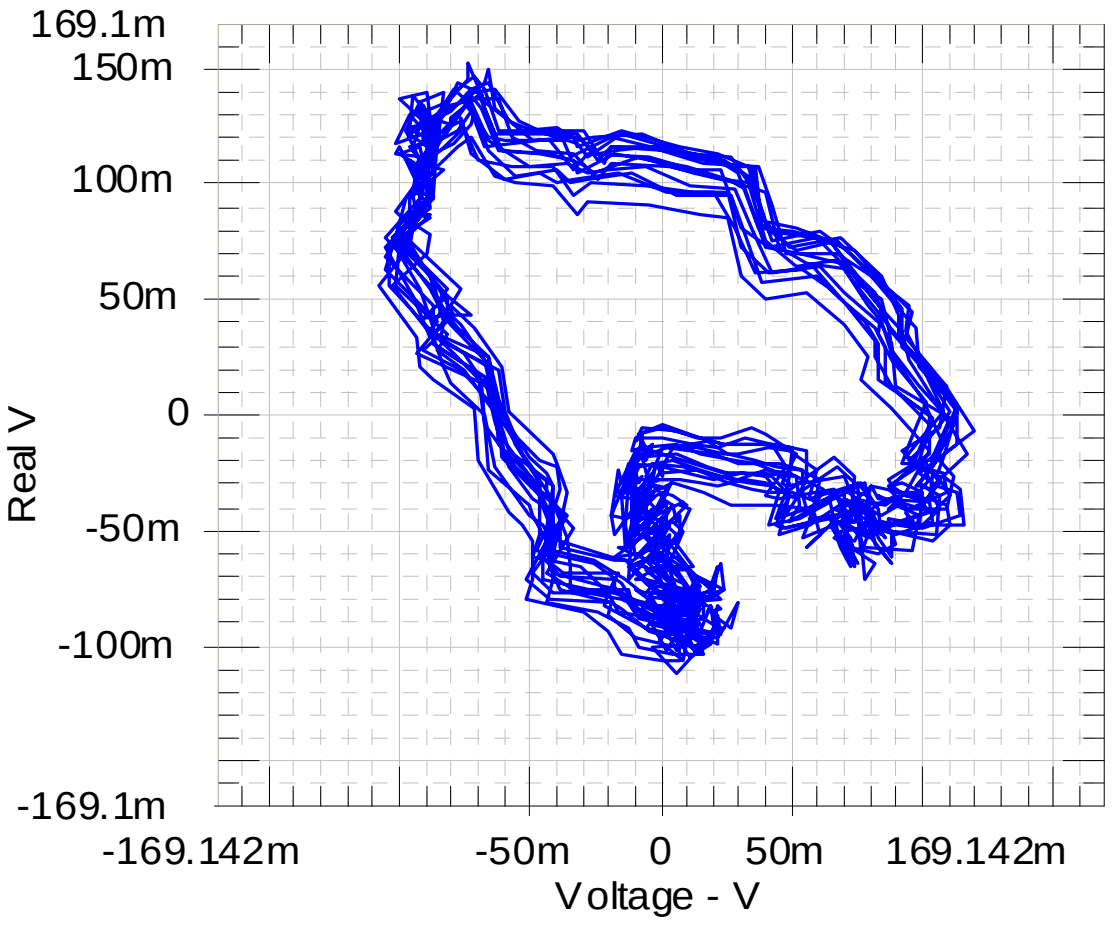

F- Without filter.

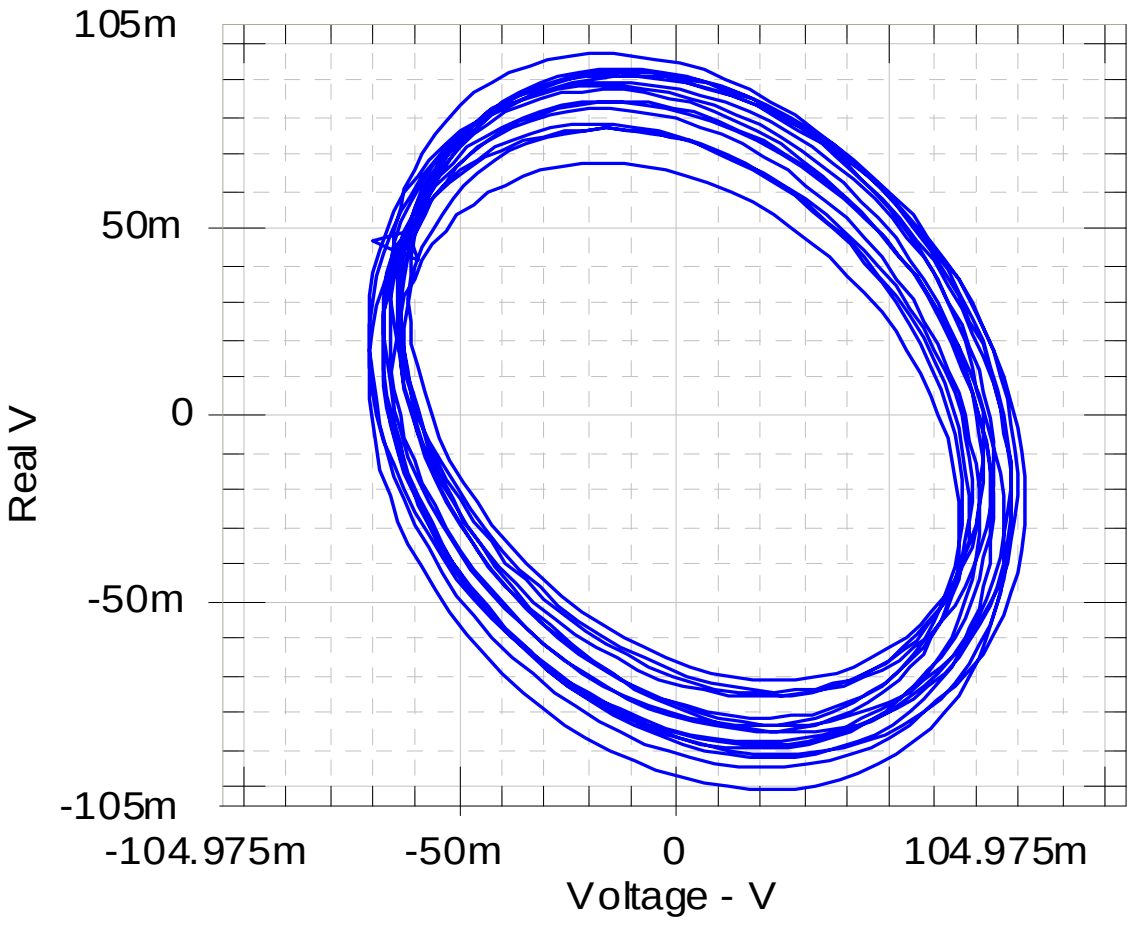

G- Without show row data filter.

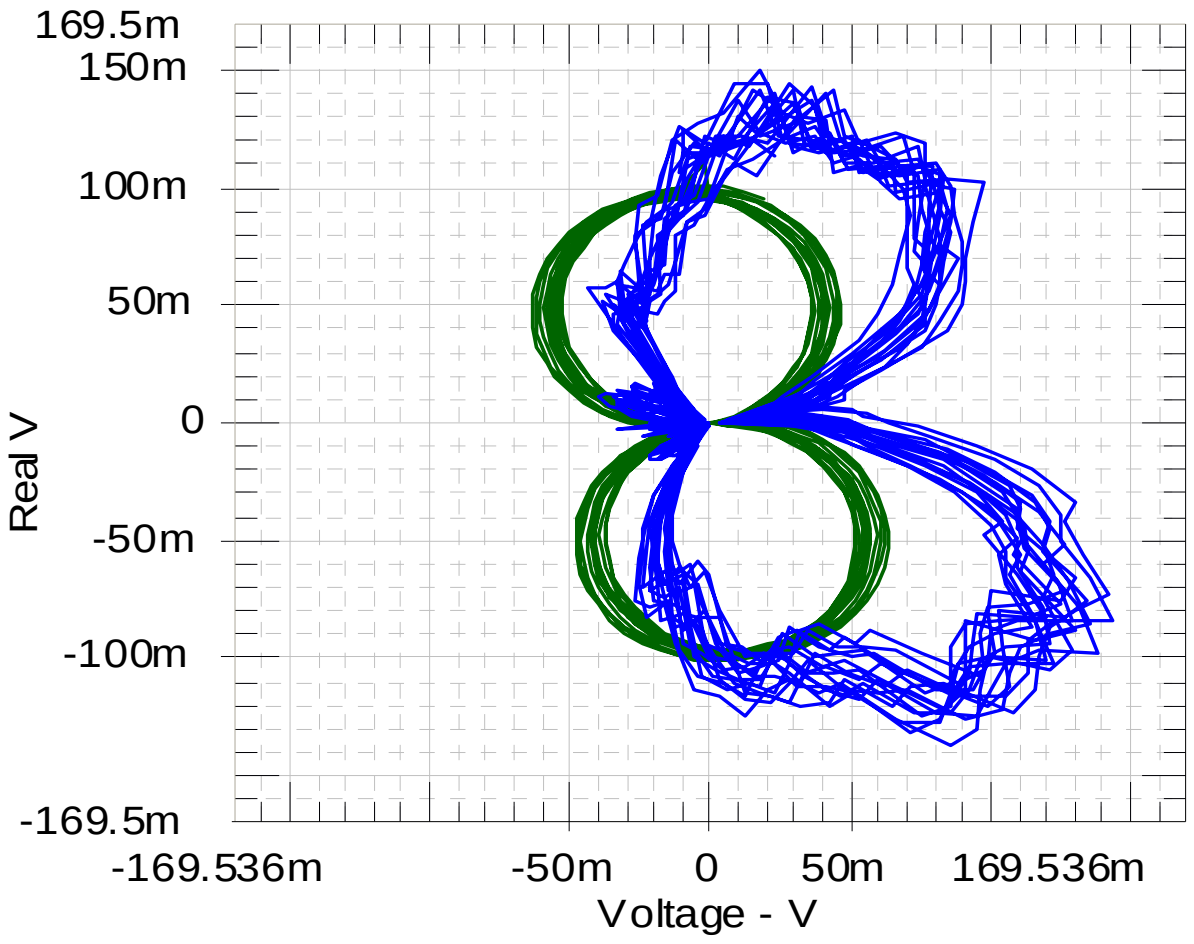

H- Plot to trigger.

Figure 3. Experimental orbit analysis fundamentals behaviour for key-phasor;

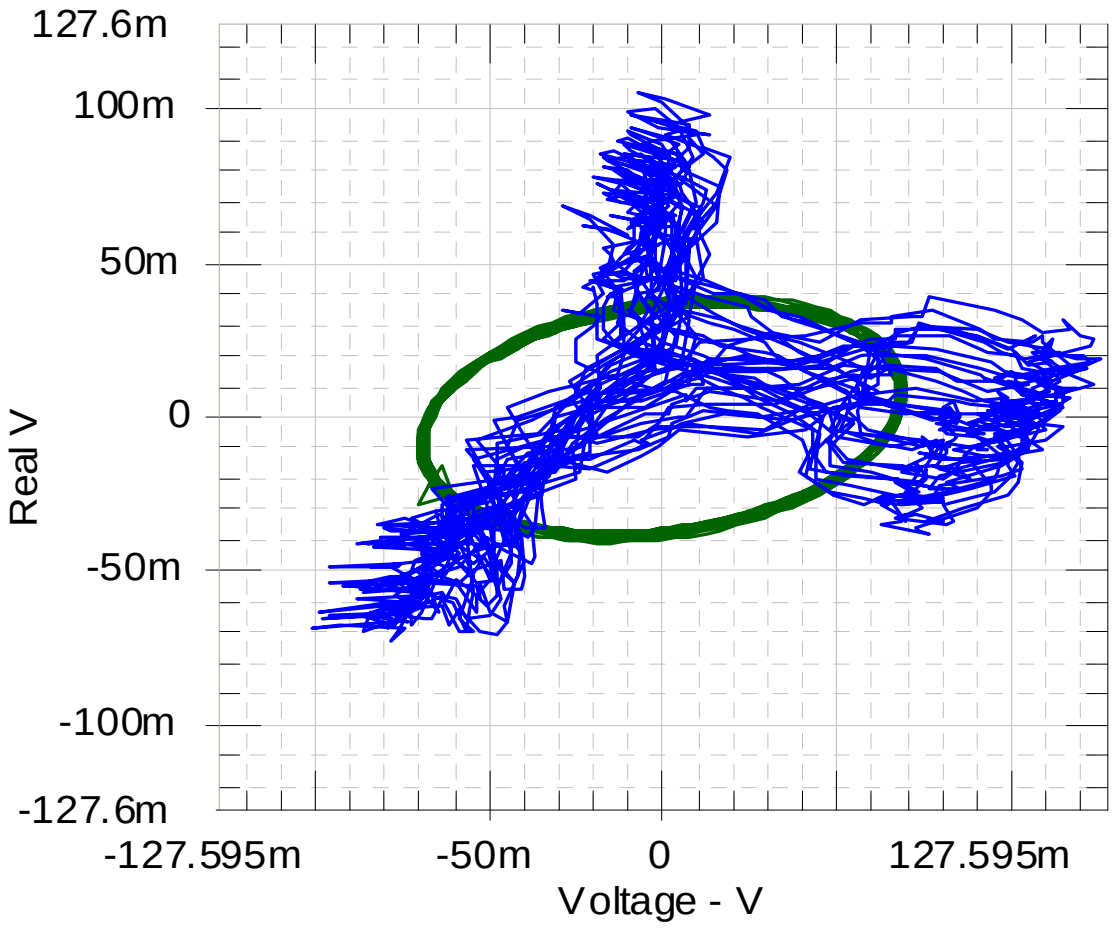

A-

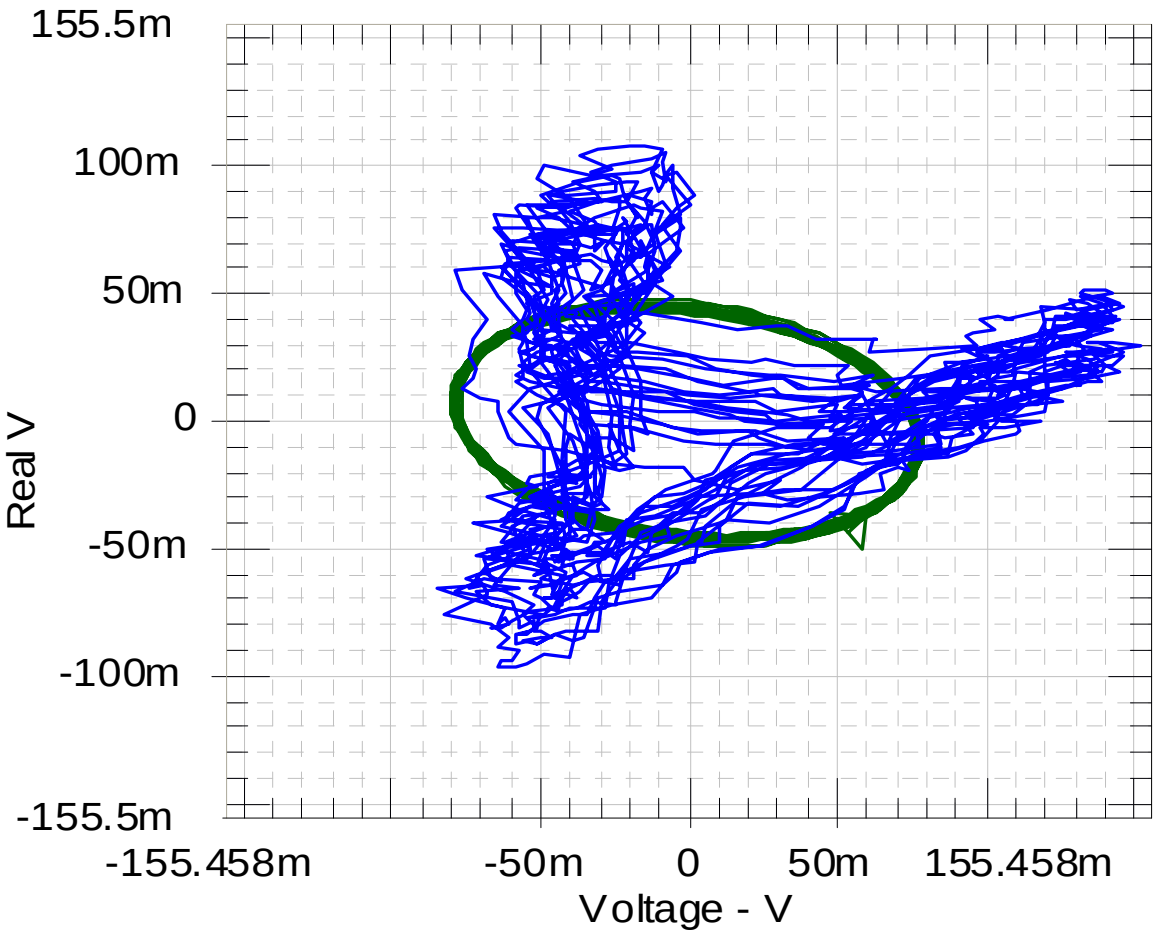

B-

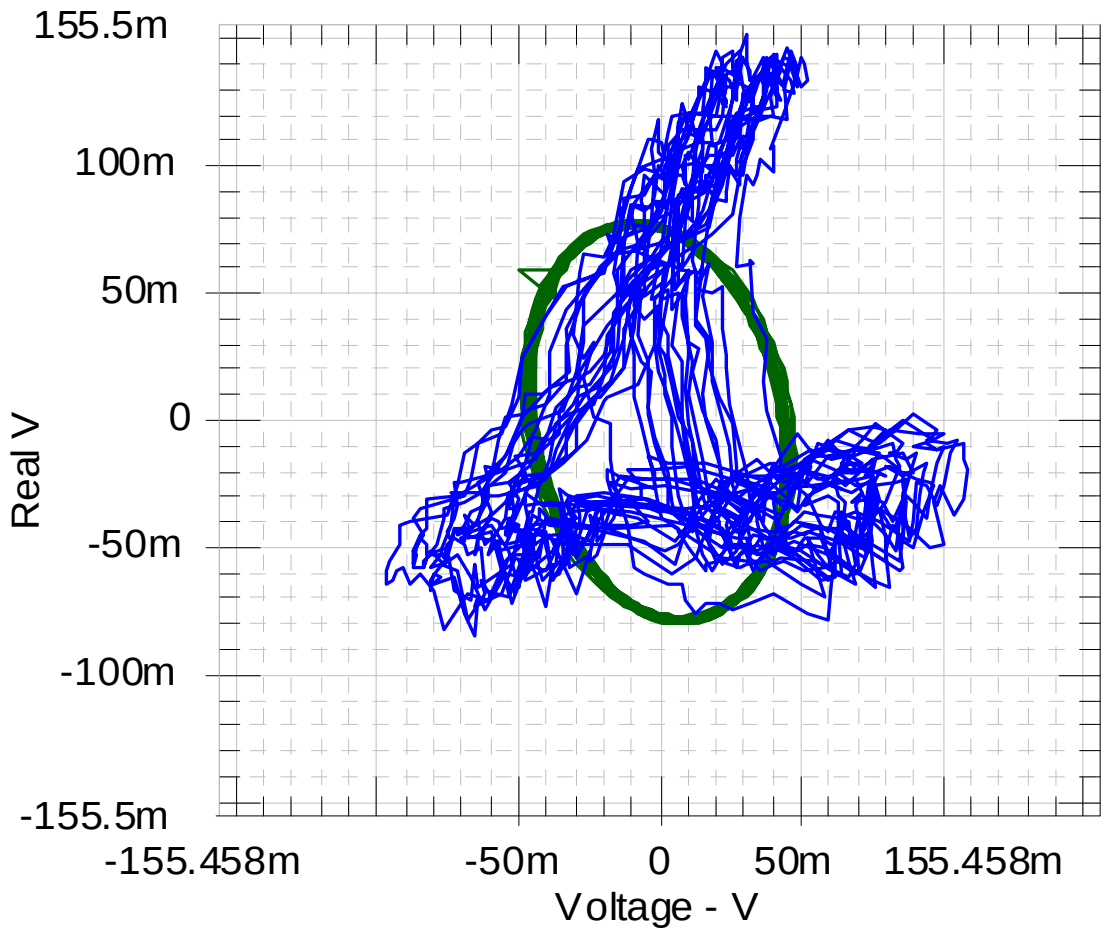

C-

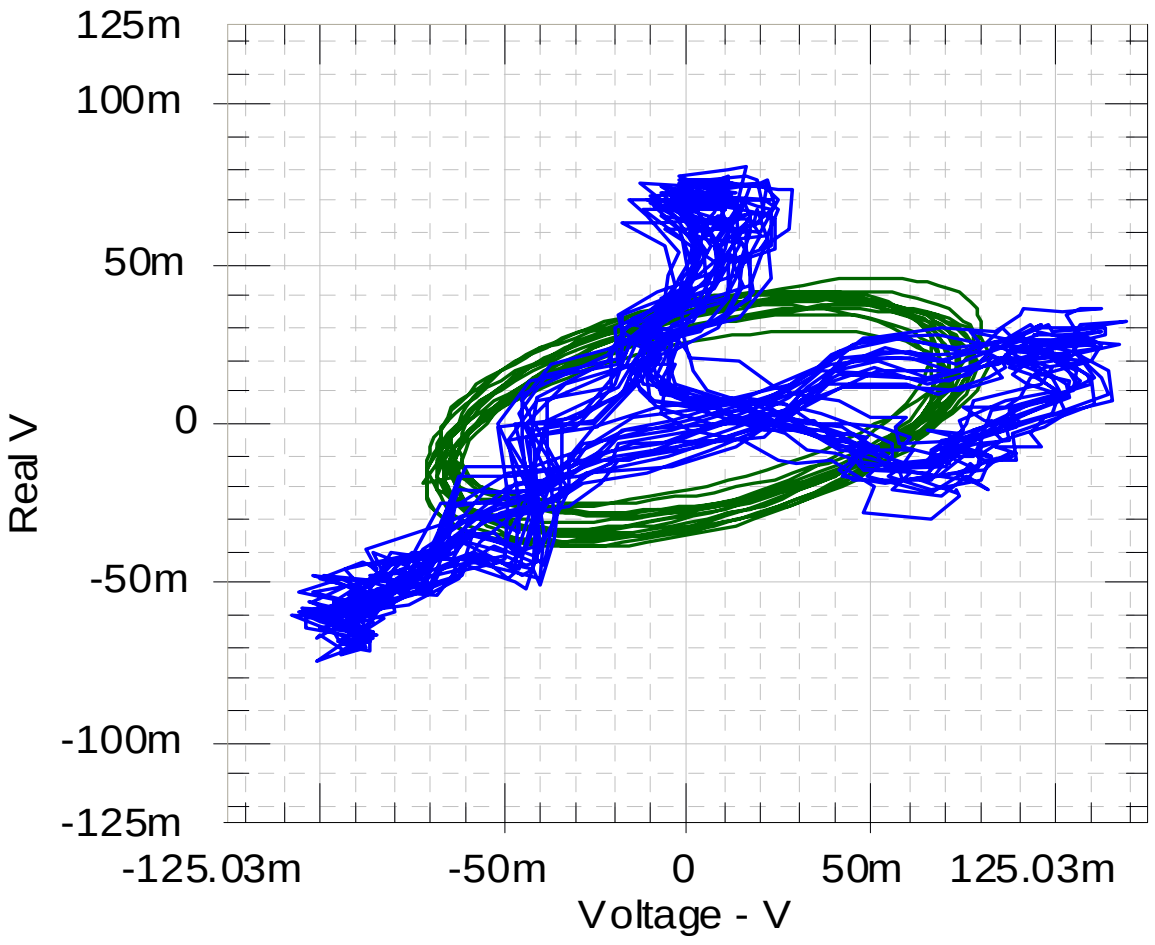

D- Without filter (HP filter).

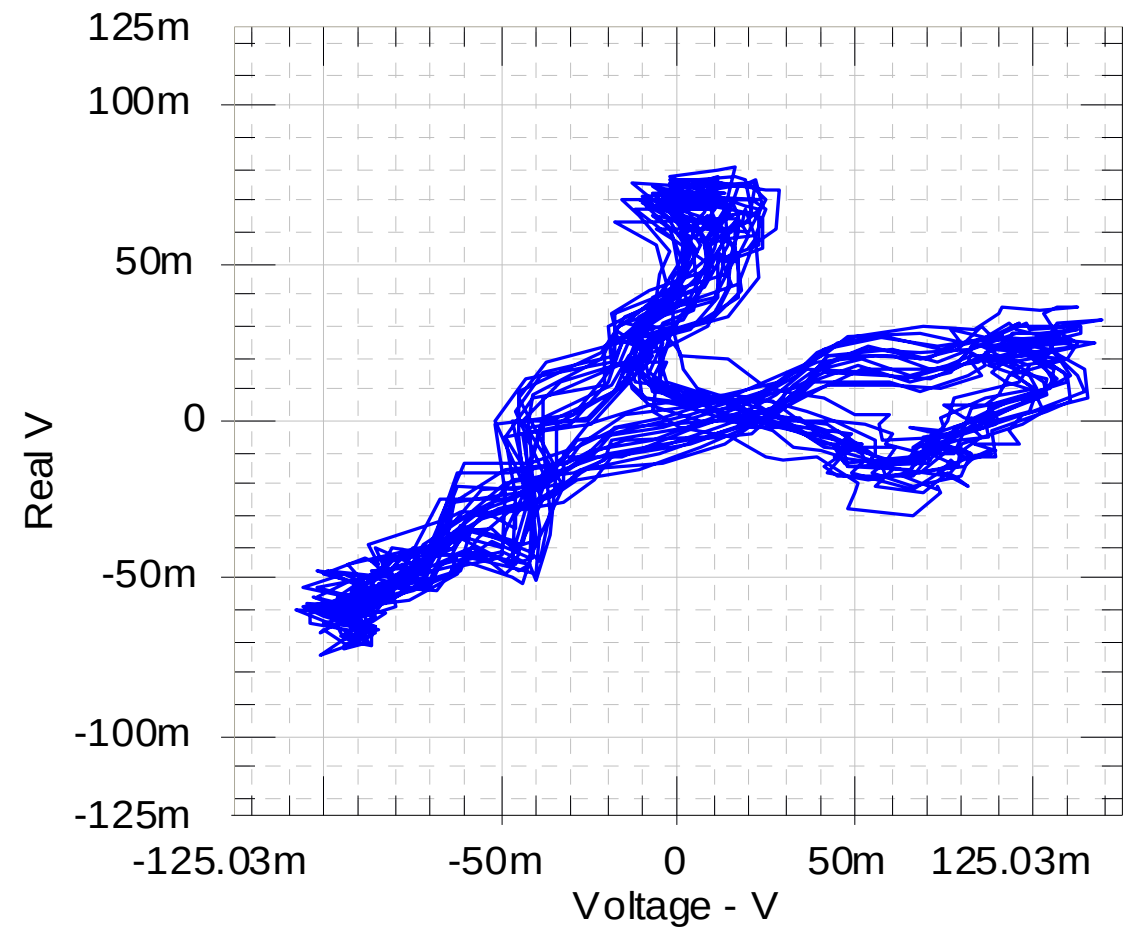

E- Without filter.

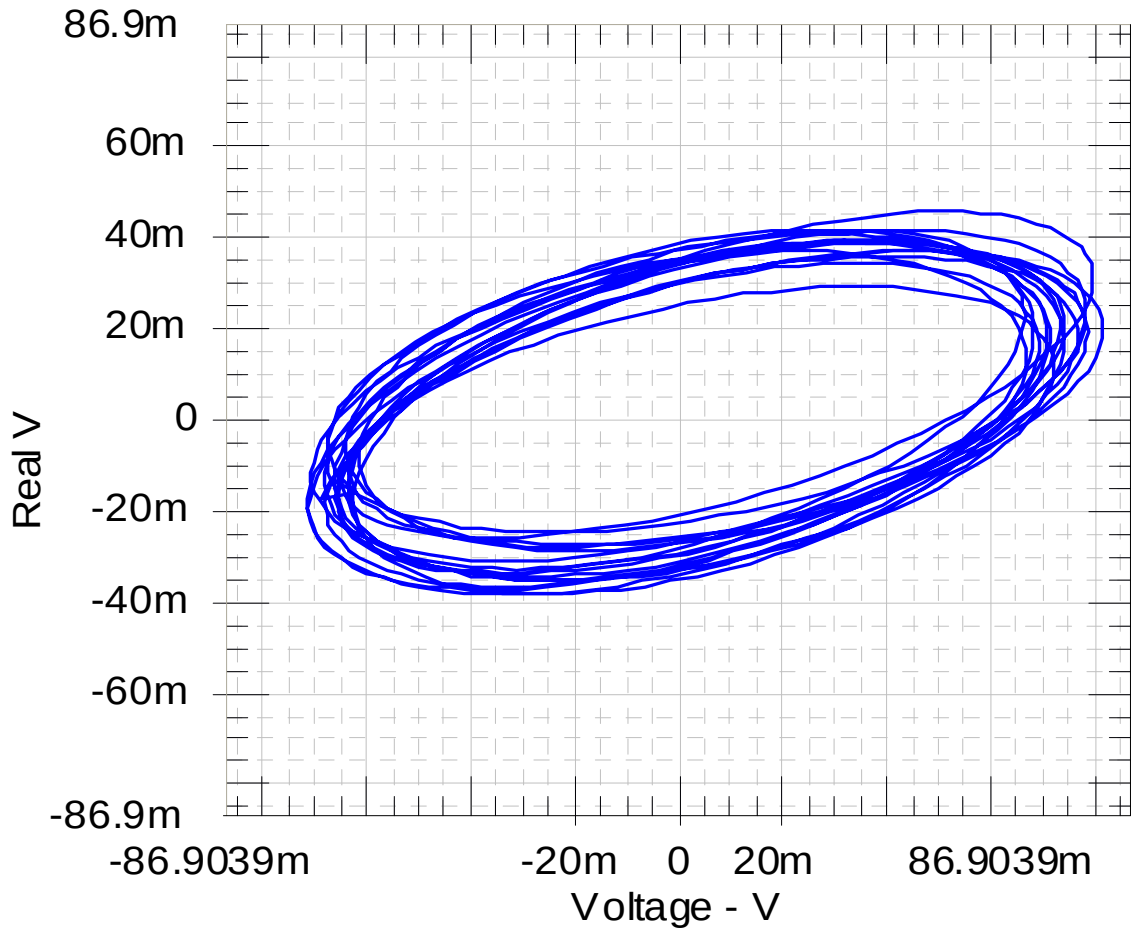

F- Without show row data filter.

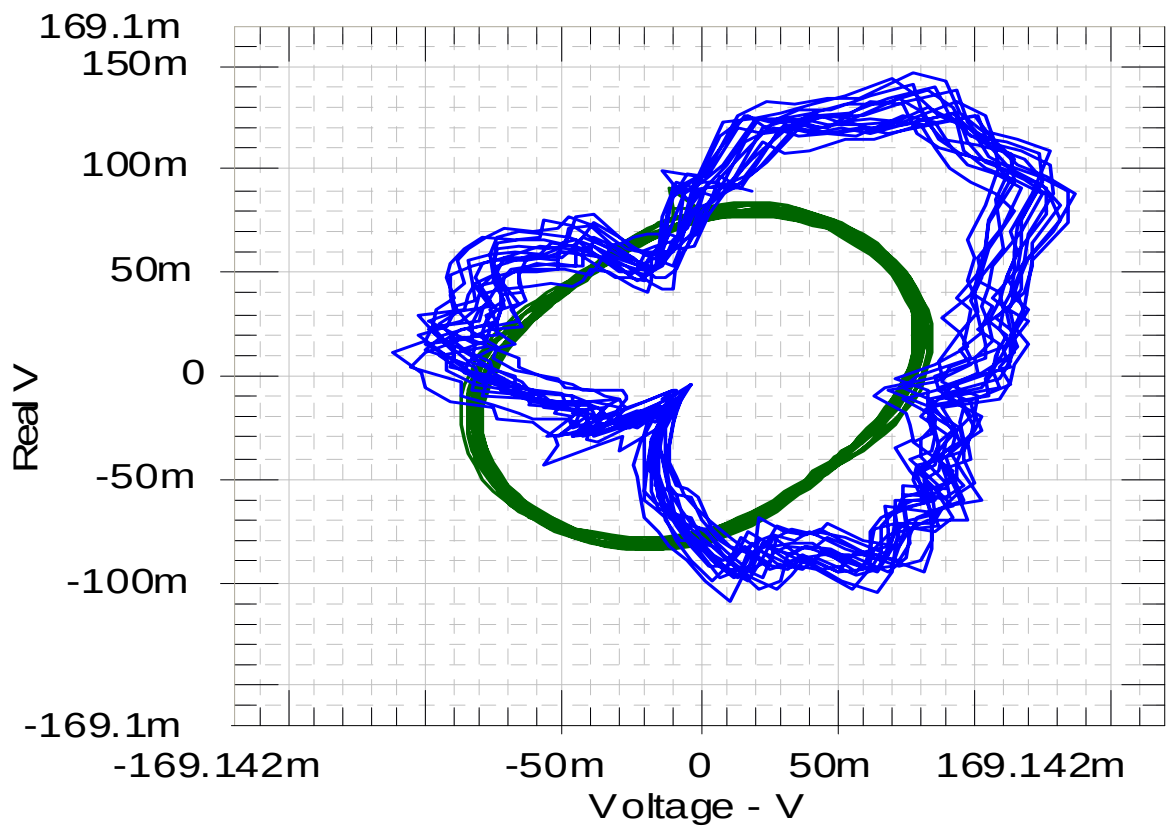

G- Plot to trigger.

Figure 4. Experimental orbits analysis fundamentals behaviour of the rotor at the different measured planes in first critical speed for channel one;

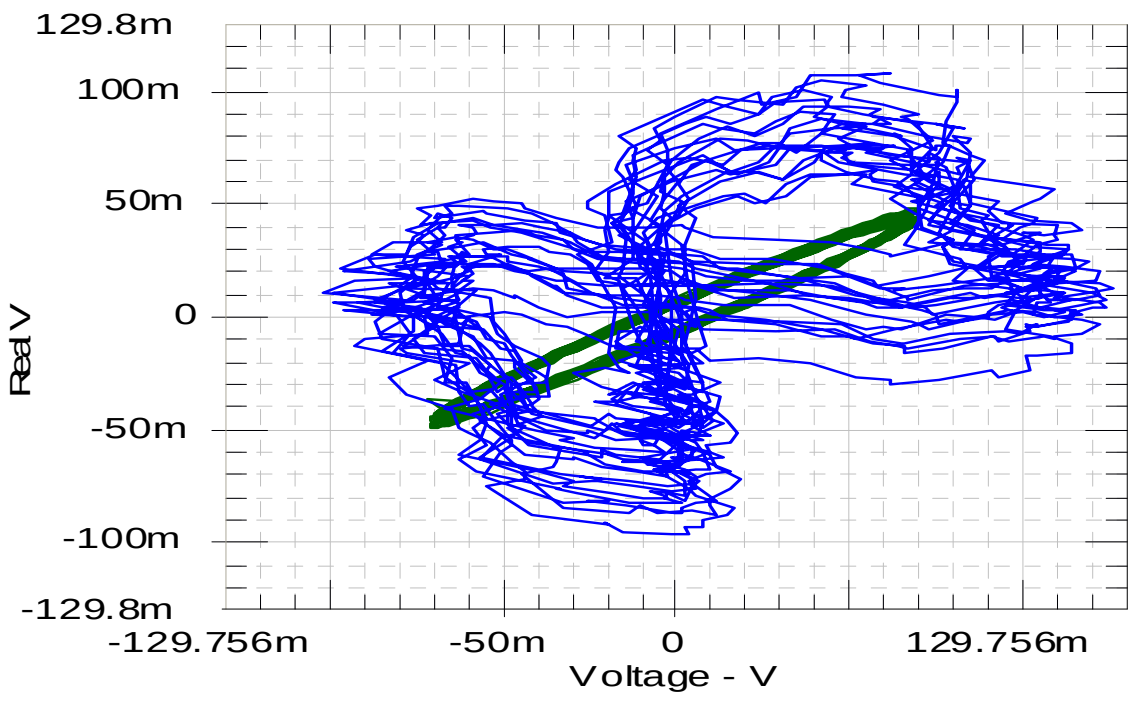

A-

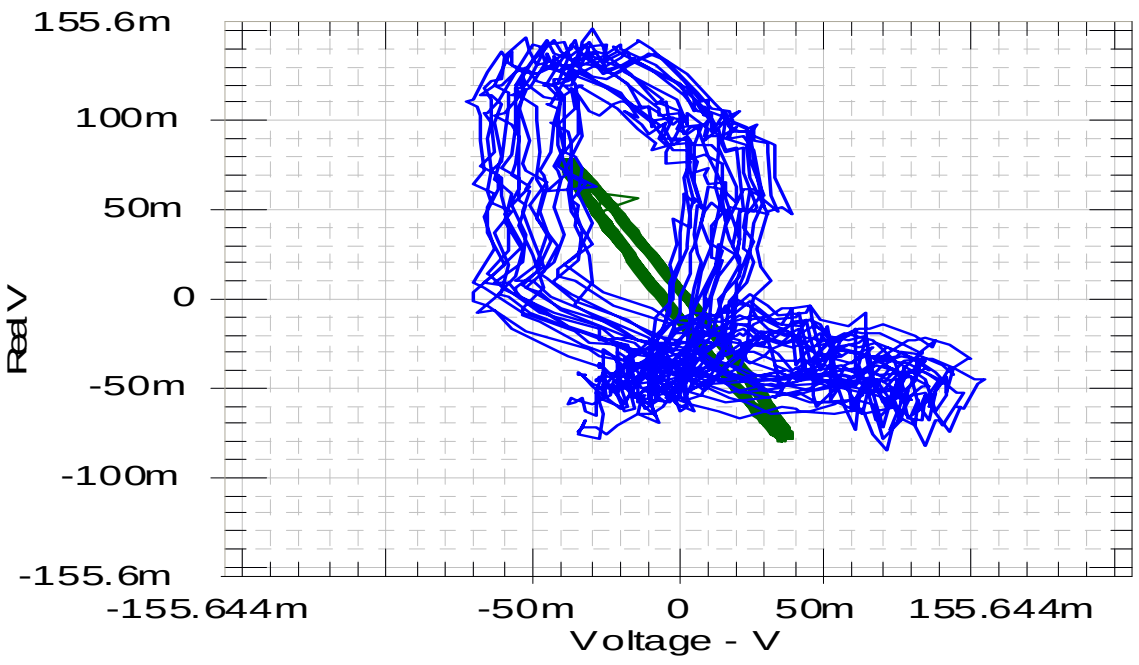

B-

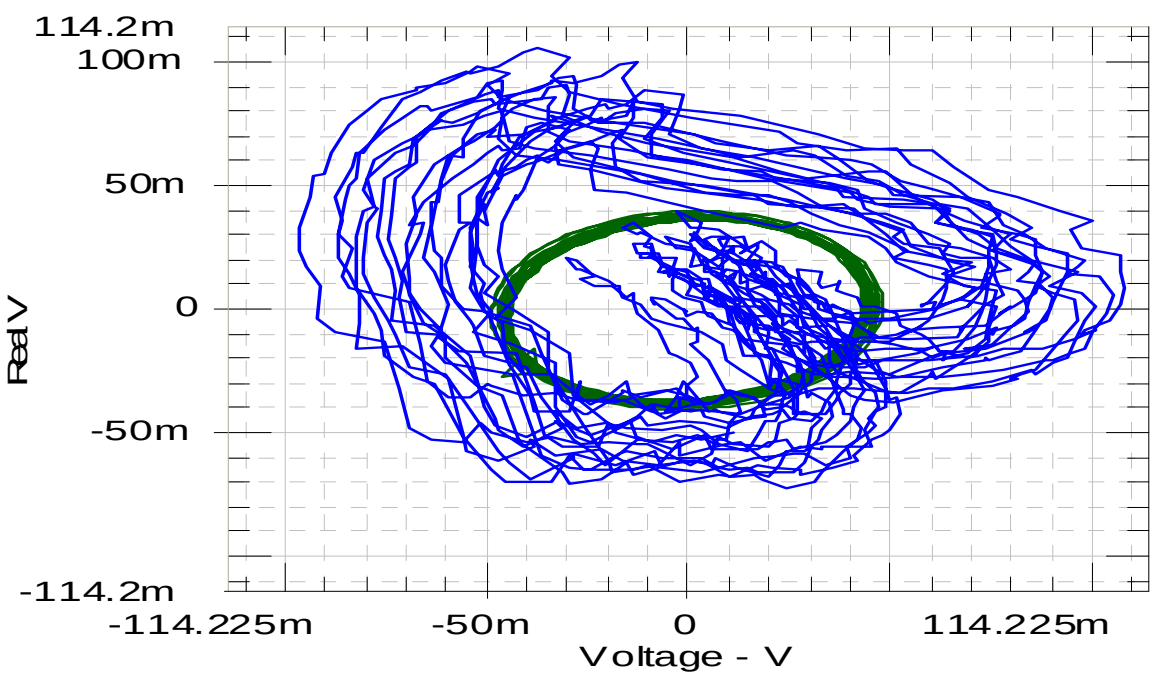

C-

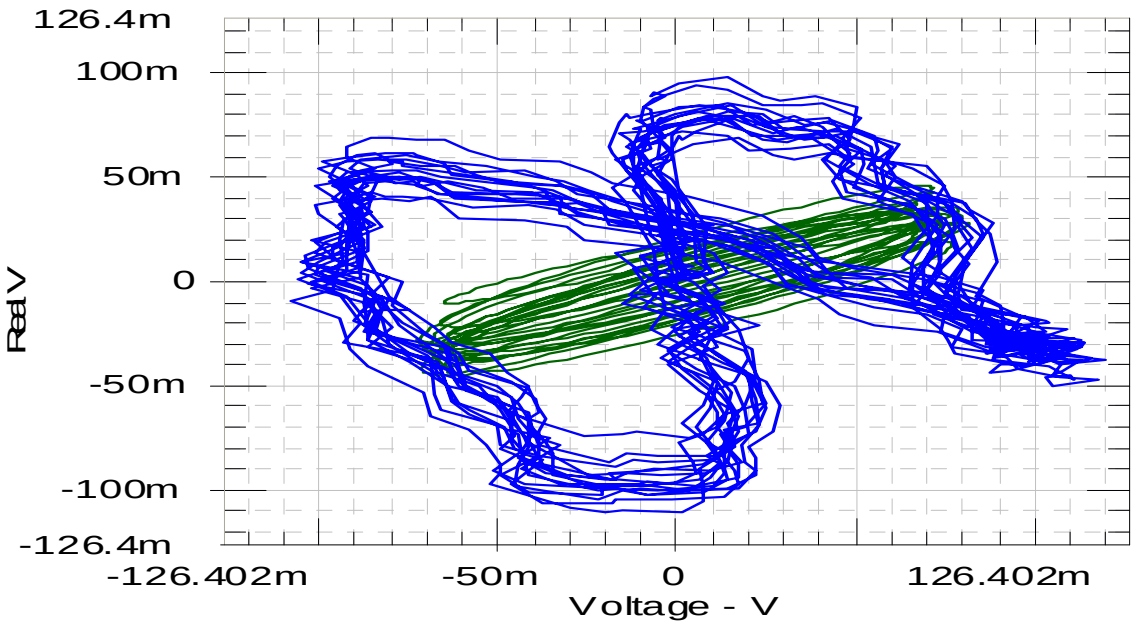

D- Without filter (HP filter).

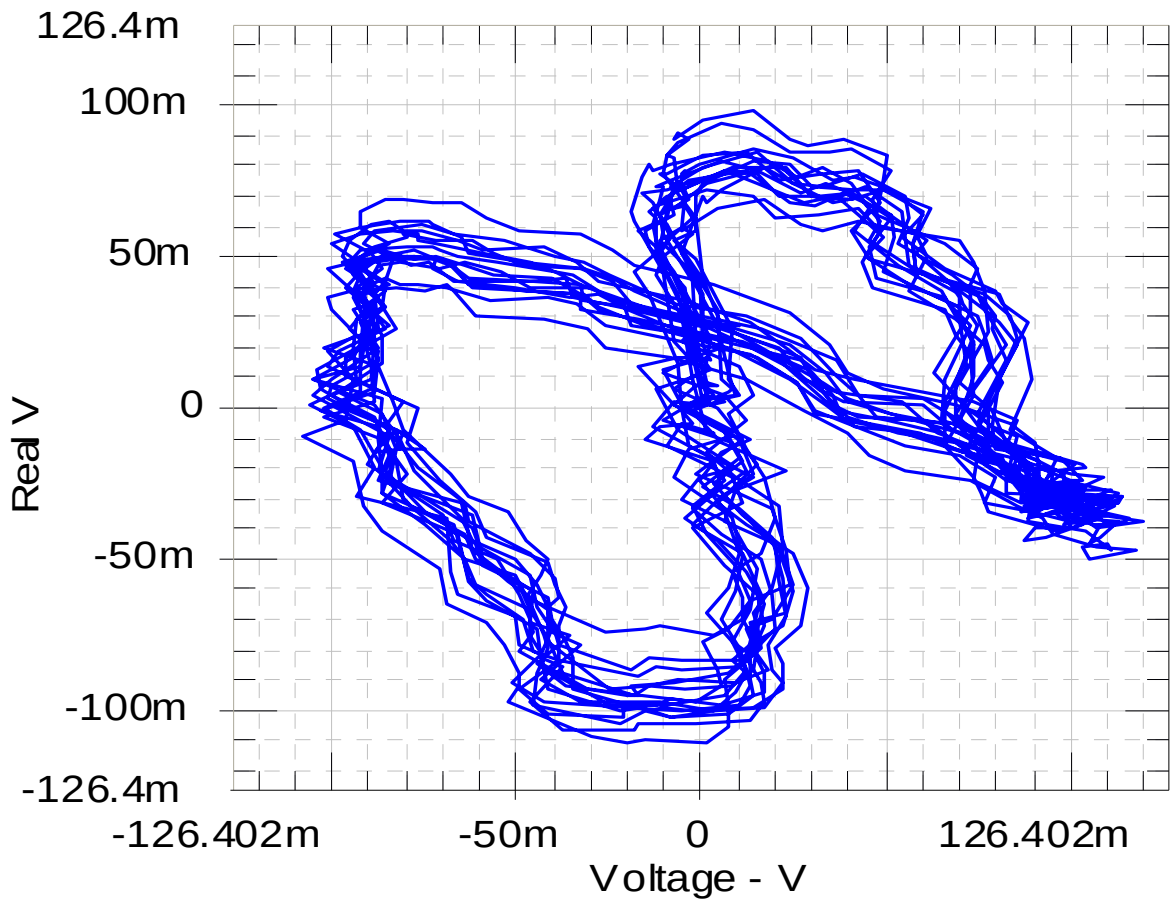

E- Without filter.

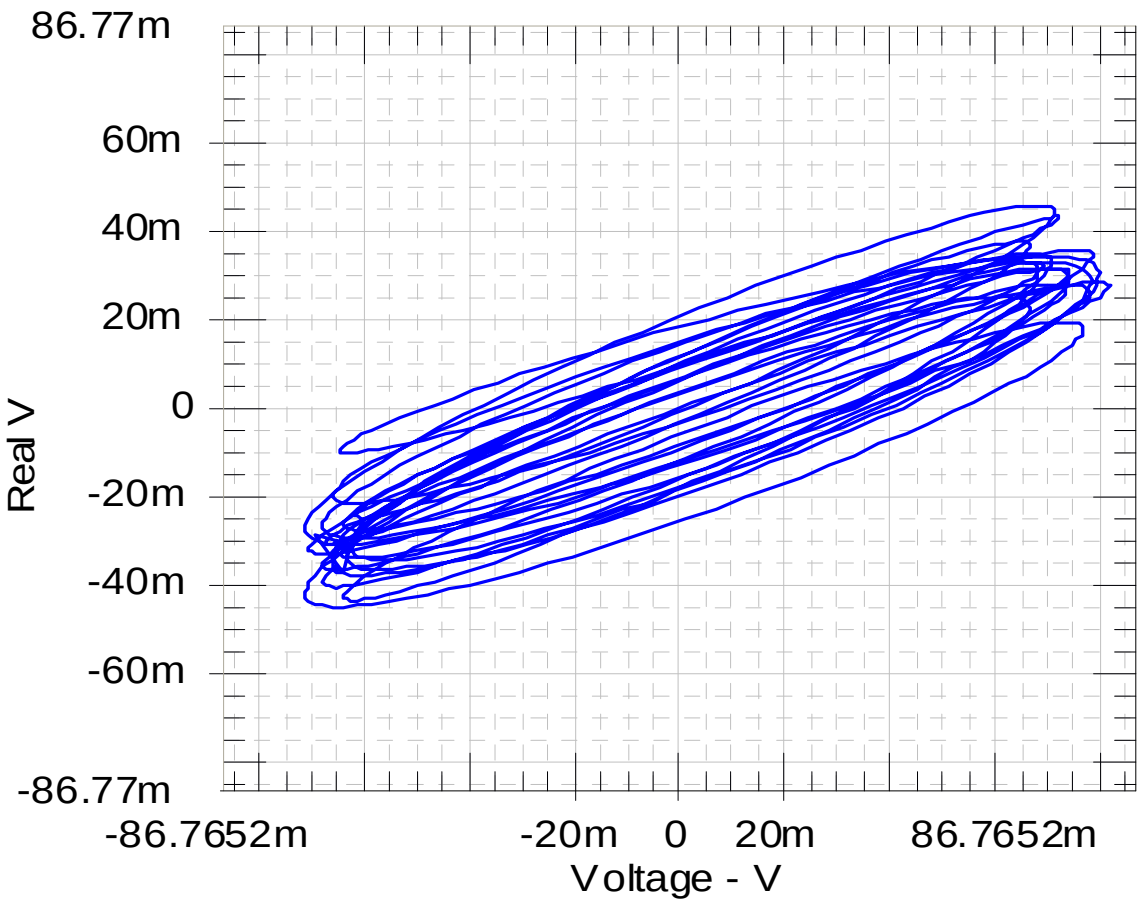

F - Without show row data filter.

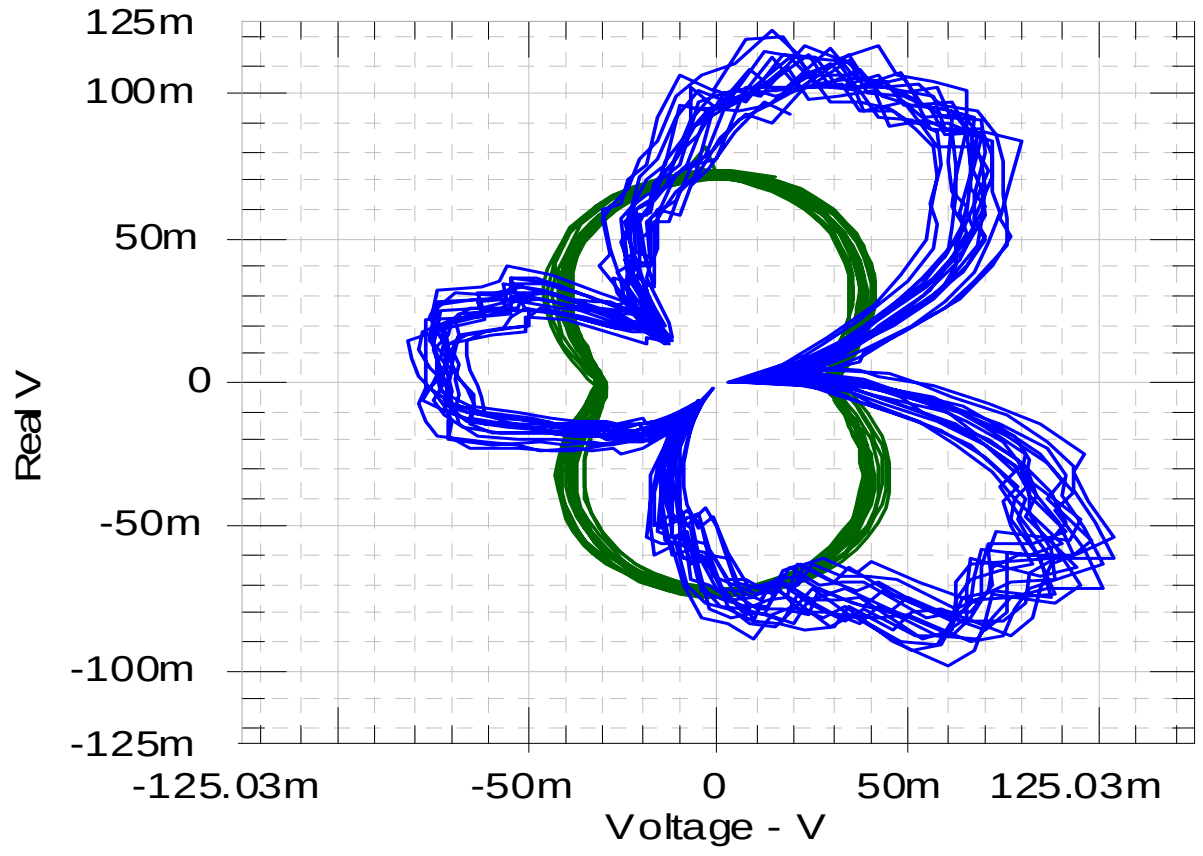

G- Plot to trigger.

Figure 5. Experimental orbits analysis fundamentals behaviour of the rotor at the different measured planes in first critical speed for channel two;

### B. *The Shock Capture Module*

The shock capture module provides shock capture, data validation and reporting. It allows you to use (m + p) library of standard test limit overlays, capture data from any number of channels, including triax's, filter the data, automatically adjust overlays for best fit [31, 32].

You will see in the display that you get an immediate readout of key pulse parameters and the whole test is controlled from one simple window for fast and efficient use even by user less familiar with the SO Analyzer and details of the application, we can see that clear in Fig.,(6).

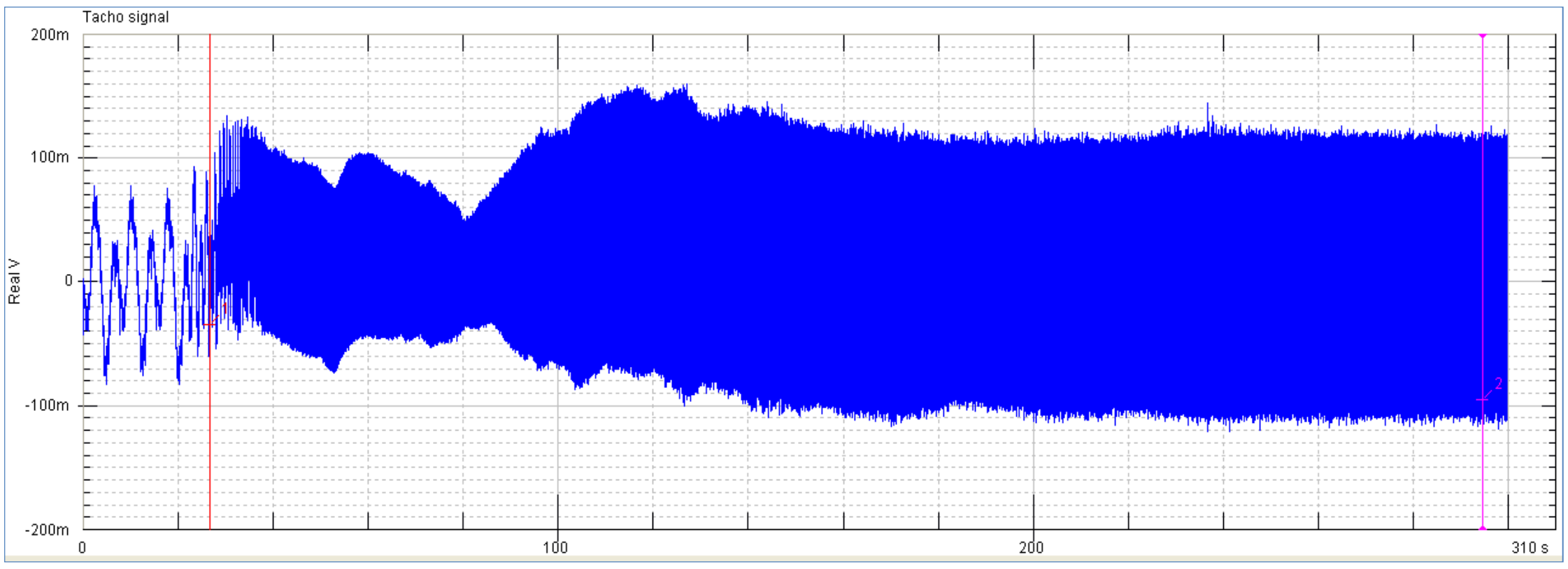

A-Tacho signal.

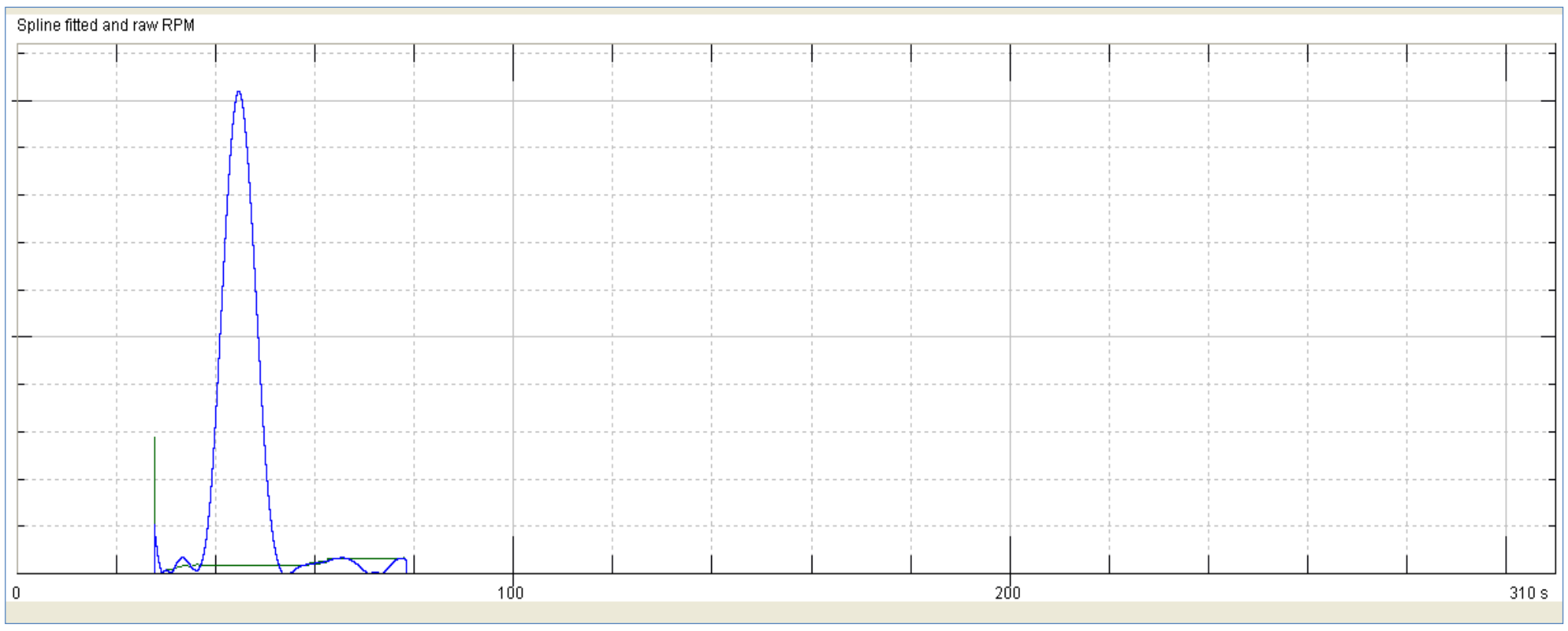

B- Before use filter.

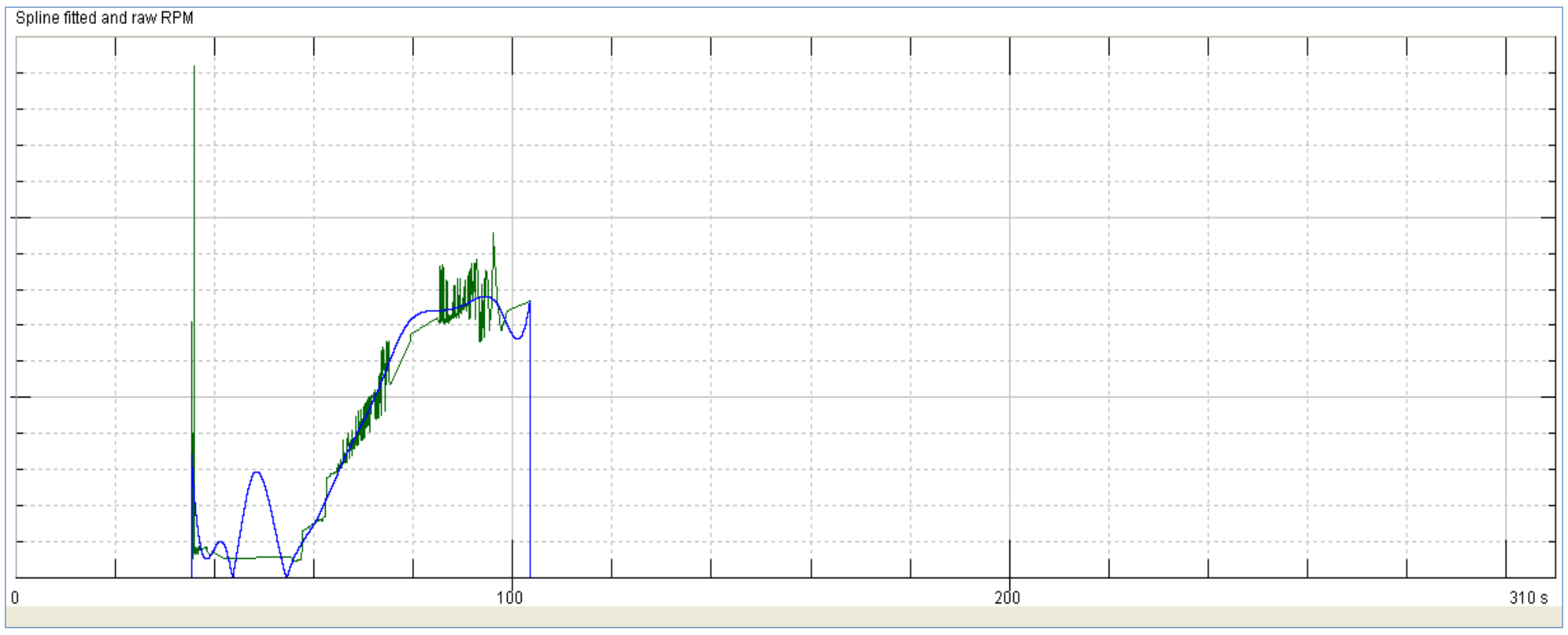

C- After used window filter with increasing the speed.

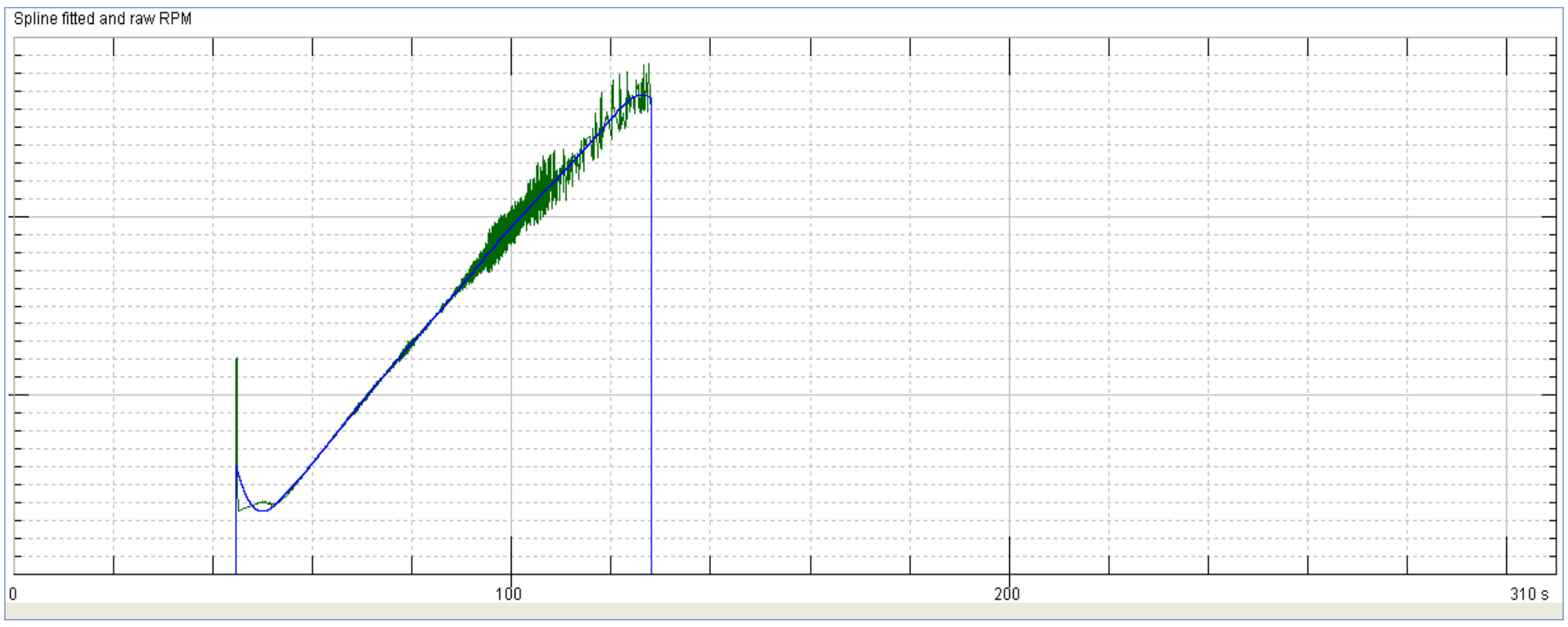

D- Improved or decay shock capture after used window filter.

Figure 6. Shock capture module;

## IV. Discussion And Conclusions

This research presented a test rig dedicated to the study of rotating machinery with the advent of computer-based data collectors and trending software. From Fig.,(2) we notes orbit analysis direction lean to left and then to the right during rotation machine because transfer the excitation harmonic force around each point in the orbit during the test, we can see that clear in Fig.,(3) demonstration accurate diagnosis of machine vibration conditions, including bearing vibration simplifies the task and increases the speed of collecting vibration monitoring data. We note From Fig.,(4) description experimental orbits in first critical speed for channel one and Fig.,(5) show orbits analysis fundamentals behaviour of the rotor system at the different measured planes in first critical speed for channel two; without filter (HP filter) see Fig.,(5-D) and without show row data filter Fig.,(5-F) shows change of performance for each point in the orbit, plot to trigger see Fig.,(5-G). From Fig.,(6) we can note shock capture module is happened at beginning when the motor turn on ( initial rotation ) this effect is very high at this region, and you can improved or decay this effect by using windowing filter that clear in Fig.,(6-D).

From the research we can conclude the vibration analysis software is an essential tool for analyzing and evaluating the mechanical vibration that occurs in rotating machinery. Vibration condition monitoring involves transmitting large quantities of data from a transducer to a separate data collector for subsequent processing and analysis. The software is then used in analysis of multiple machine vibration parameters to assess operating performance. Maintenance personnel then use the data to determine if unscheduled maintenance is necessary, or if shutdown is required. Compact portable vibration data collectors are easy to use with a portable computer, such as a laptop or note book multiple channel models are available, usually do not require any special training.

## V. Acknowledgment

The authors are deeply appreciative to (SEC) Faculty of Science, Engineering and Computing in Kingston University London that provides technical support for the research, and the Iraqi Ministry of Higher Education, Iraqi Cultural Attaché in London for that provides funding for the research.

## NOMENCLATURE

| | |
|---|---|
| [M] | Mass matrix, speed dependent |
| [C] | Damping matrix, speed dependent |
| [K] | Stiffness matrix, speed dependent |
| [G] | Gyroscopic matrix of rotating system |
| {F} | Force vector |
| $\Omega$ | Rotating speed |
| $\varphi_{jr}$ | Normalised Eigen vector |
| $\varphi_{kr}$ | Normalised Eigen vector |
| $\alpha_{jk}$ | The receptance for one measurement between two coordinates j and k |
| Im | Imaginary |
| Re | Reynolds number |
| $\zeta_r$ | Damping ratio of the rth mode |
| $\omega, \omega_r$ | Excitation, Natural frequency of the rth mode (modal parameters) |
| q , r | Are generally two different modes |
| t | Time variable |
| n | Degree of freedom/coordinates |

## AUTHOR'S INFORMATION

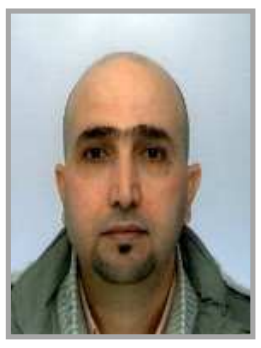

Mr. Hisham A. H. Al-Khazali1, He has PhD Student in Kingston University London, and (SEM) member, Society for Experimental Mechanics. Inc., in USA. He was born in 28 Aug 1973 Baghdad/Iraq. He received his BSc (Eng.), in Mechanical Engineering (1996), University of Technology, Baghdad. MSc in Applied Mechanics, University of Technology, Baghdad (2000).
E-mail, k0903888@kingston.ac.uk

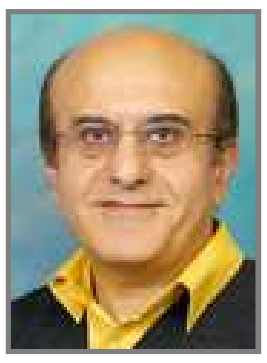

Dr. Mohamad R. Askari2, BSc (Eng), MSc, PhD, CEng, MIMechE, MRAeS. He has (Principal Lecturer, Blended Learning Coordinator), Member teaching staff in Kingston University London, His Teaching Area: Applied Mechanics, Aerospace Dynamics, Dynamics and Control, Structural and Flight Dynamics, Engineering Design, Software Engineering to BEng Mechanical and Aerospace second and final years. He was Year Tutor for BEng, Mechanical Engineering Course and School Safety Advisor.
E-mail, M.Askari@kingston.ac.uk